\documentclass[sigconf,authorversion,nonacm]{acmart} % for arxiv

\AtBeginDocument{%
  \providecommand\BibTeX{{%
    \normalfont B\kern-0.5em{\scshape i\kern-0.25em b}\kern-0.8em\TeX}}}

\usepackage{color}
\usepackage{enumitem}
\usepackage[utf8]{inputenc}
\usepackage{graphicx,dblfloatfix}
\usepackage{multirow}
\usepackage[title]{appendix}
\usepackage{ifthen}

\begin{document}

\title{Automatic Multi-Path Web Story Creation from a Structural Article}
% \protect\\

% \acmConference[]{Submitted to MobileHCI'23}{2023}{}
% \acmPrice{15.00} 
% \acmISBN{978-1-4503-XXXX-X/18/06}

%\author{Daniel Nkemelu, Peggy Chi, Daniel Castro Chin, Krishna Srinivasan, Irfan Essa}
\author{Daniel Nkemelu}
\authornote{Work was done during an internship at Google.}
\affiliation{%
  \institution{Georgia Institute of Technology}
  \city{Atlanta, GA}
  \country{USA}
}

\author{Peggy Chi}
\affiliation{%
  \institution{Google Research}
  \city{Mountain View, CA}
  \country{USA}
}

\author{Daniel Castro Chin}
\affiliation{%
  \institution{Google Research}
  \city{Atlanta, GA}
  \country{USA}
}

\author{Krishna Srinivasan}
\affiliation{%
  \institution{Google Research}
  \city{Mountain View, CA}
  \country{USA}
}

\author{Irfan Essa}
\affiliation{%
  \institution{Google Research}
  \institution{Georgia Institute of Technology}
  \city{Atlanta, GA}
  \country{USA}
}

%%
%% The "author" command and its associated commands are used to define
%% the authors and their affiliations.
%% Of note is the shared affiliation of the first two authors, and the
%% "authornote" and "authornotemark" commands
%% used to denote shared contribution to the research.
% \author{Ben Trovato}
% \authornote{Both authors contributed equally to this research.}
% \email{trovato@corporation.com}
% \orcid{1234-5678-9012}
% \author{G.K.M. Tobin}
% \authornotemark[1]
% \email{webmaster@marysville-ohio.com}
% \affiliation{%
%   \institution{Institute for Clarity in Documentation}
%   \streetaddress{P.O. Box 1212}
%   \city{Dublin}
%   \state{Ohio}
%   \country{USA}
%   \postcode{43017-6221}
% }

\renewcommand{\shortauthors}{Nkemelu et al.}
\newcommand{\systemname}{Wiki2Story}
\def\systemname{Wiki2Story}
\def\numpagesrun{500}
\def\etal{et al.}
\newcommand{\userquote}[1]{``\emph{#1}''}
\newcommand{\subheading}[1]{\textbf{#1}}
\newcommand{\principle}[1]{\textbf{#1}}
\newcommand{\code}[1]{\texttt{\small\textcolor{blue}{#1}}}

\newcommand{\mainexample}{Apple}
\newcommand{\mainexampleurl}{https://en.wikipedia.org/wiki/Apple}

\def\etal{et al.}
\newcommand{\note}[1] {\textcolor{red}{[NOTE:#1]}}
\newcommand{\irfan}[1] {\textcolor{purple}{[IE:#1]}}

\newcommand{\added}[1] {\textcolor{black}{#1}}
\newcommand{\mobilehci}[1] {\textcolor{black}{#1}}

%%
%% The abstract is a short summary of the work to be presented in the
%% article.
\begin{abstract}
Web articles such as Wikipedia serve as one of the major sources of knowledge dissemination and online learning. However, their in-depth information--often in a dense text format--may not be suitable for mobile browsing\mobilehci{, even in a responsive UI}. We propose an automatic approach that converts a structural article of any length into a set of interactive Web Stories that are ideal for mobile experiences. We focused on Wikipedia articles and developed \mobilehci{Wiki2Story, a pipeline based on language and layout models}, to demonstrate the concept. Wiki2Story dynamically slices an article and plans one to multiple Story paths according to the document hierarchy. For each slice, it generates a multi-page summary Story \mobilehci{composed of text and image pairs in visually-appealing layouts}. We derived design principles from an analysis of manually-created Story practices. We executed our pipeline on 500 Wikipedia documents and conducted user studies to review selected outputs. Results showed that Wiki2Story effectively captured and presented salient content from the original articles and sparked interest in viewers. % for readers to review digestible content.
% Wikipedia serves as a global learning community of users mutually-engaged in knowledge production and sharing. However, dense text content as found in many Wikipedia articles can be difficult to follow, understand, and recall compared to multimedia forms of presentation. A growing percentage of internet users also choose or need multimedia content as a preferred means of learning new material. We present Wiki2Story, an automatic approach that converts a Wikipedia page of any length into an interactive Accelerated Mobile Pages (AMP) story format. Using a state-of-the-art text summarization model, WIki2Story’s engine performs section-by-section abstractive summarization of a wikipedia page, creates multiple story navigation paths by traversing a tree structure of the Wikipedia article, and dynamically selects relevant text-image layouts for presenting summarized content to the user. Our user evaluation shows that Wiki2Story effectively captures and presents salient content from the original article and [is a preferred means of learning about new Wikipedia articles]. Wiki2Story promises a scalable cost-efficient approach for generating accessible, digestible, and interactive versions of massive text content available online.
\end{abstract}

%%
%% The code below is generated by the tool at http://dl.acm.org/ccs.cfm.
%% Please copy and paste the code instead of the example below.
%%
\begin{CCSXML}
<ccs2012>
   <concept>
       <concept_id>10003120.10003121</concept_id>
       <concept_desc>Human-centered computing~Human computer interaction (HCI)</concept_desc>
       <concept_significance>500</concept_significance>
       </concept>
 </ccs2012>
\end{CCSXML}

\ccsdesc[500]{Human-centered computing~Human computer interaction (HCI)}

\keywords{Story Format, Text Summarization, Wikipedia, Web Story, Slideshow}

\begin{teaserfigure}
  \centering
  \includegraphics[width=\textwidth]{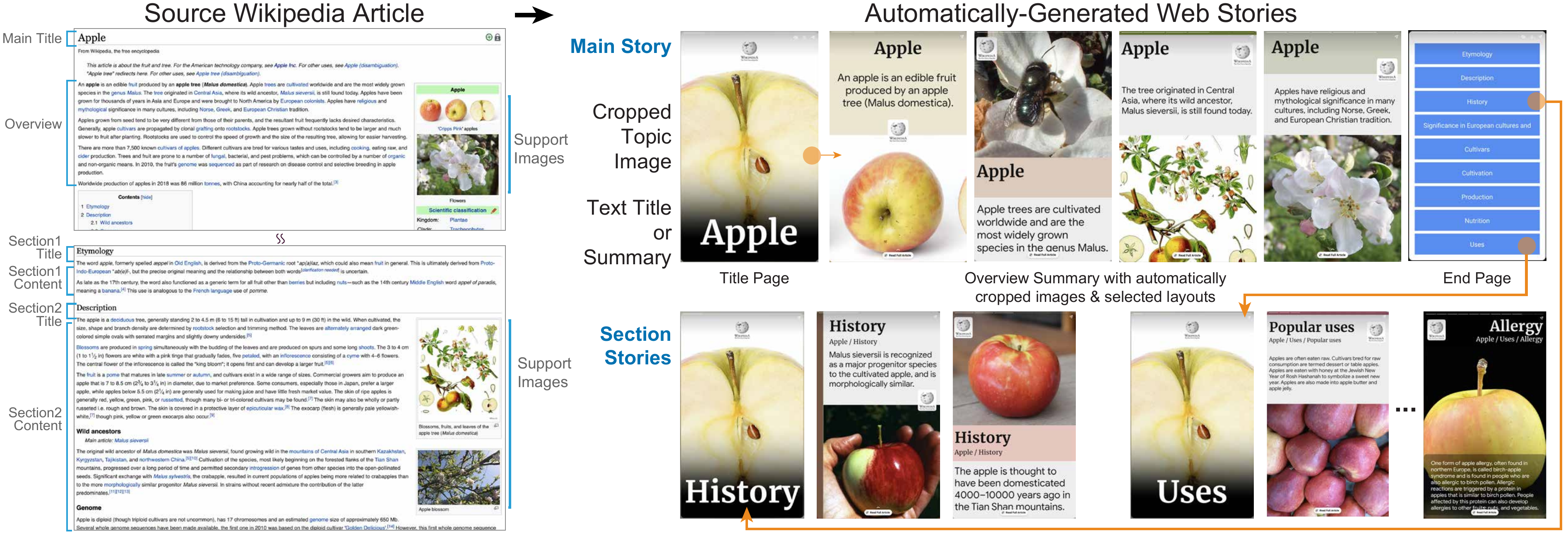}
  \Description[]{}
  \caption{\systemname{} automatically converts a structural \mobilehci{article into a set of Web Stories for mobile browsing. It summarizes the article sections, matches image to text summaries based on image features, and renders as consecutive Story pages, each with a suitable graphical layout. Users can tap through the pages and navigate to the subsection Stories}. They can follow the ``Read Full Article'' link to view the original section in the source article.
  \emph{Source article: ``\href{https://en.wikipedia.org/wiki/Apple}{\mainexample{}}'' by Wikipedia is licensed under CC-By SA 3.0.}
  % \emph{Source document: ``\href{https://en.wikipedia.org/wiki/African_elephant}{African elephant}'' by Wikipedia is licensed under CC-By SA 3.0.}
  }
  \label{fig:teaser}
\end{teaserfigure}

%Figure 1: Wiki2Story automatically converts a structural text article into a set of Web Stories based on the document hierarchy.It summarizes the text paragraphs, locates relevant images from the article, and renders into multiple lists of Story pages, eachwith a suitable layout that presents cropped images and a text summary. Users can tap through the pages to review and navigateto the subsections, each as a Web Story. They can follow the “Read Full Article” link to view the source article section.Sourcearticle: “Apple” by Wikipedia is licensed under CC-By SA 3.0
\maketitle

\section{Introduction}
Online articles help users learn about diverse topics and engage with new ideas. One major platform serving such articles is Wikipedia, forming a global community for knowledge sharing~\cite{Wikipedia_CHI18,Wikipedia_CHI20,bruckman2022should}. As of \mobilehci{May 2023, the English version of Wikipedia included over 6.6 million articles, with an average of 552} new articles produced daily~\cite{WikiStats}.
Such articles often contain in-depth content about a topic, including a background, multiple hierarchical levels of context, supporting visuals, and references.
\mobilehci{They are conventionally ideal for consumption with a large screen such as a desktop or laptop, rather than for mobile viewing. Thanks to responsive web design, with proper annotations, mobile browsers can automatically organize content into flexible layouts adapted to varying screen sizes and orientations~\cite{gardner2011responsive, mohorovivcic2013implementing}. This requires authors to manually annotate the HTML elements to prioritize or hide content via the frameworks. For long articles edited by a community, such as Wikipedia, it can be challenging for authors to properly annotate \emph{with a consideration to optimize the viewing user experiences.}}

\begin{figure*}[t!]
  \centering
  \includegraphics[width=\textwidth]{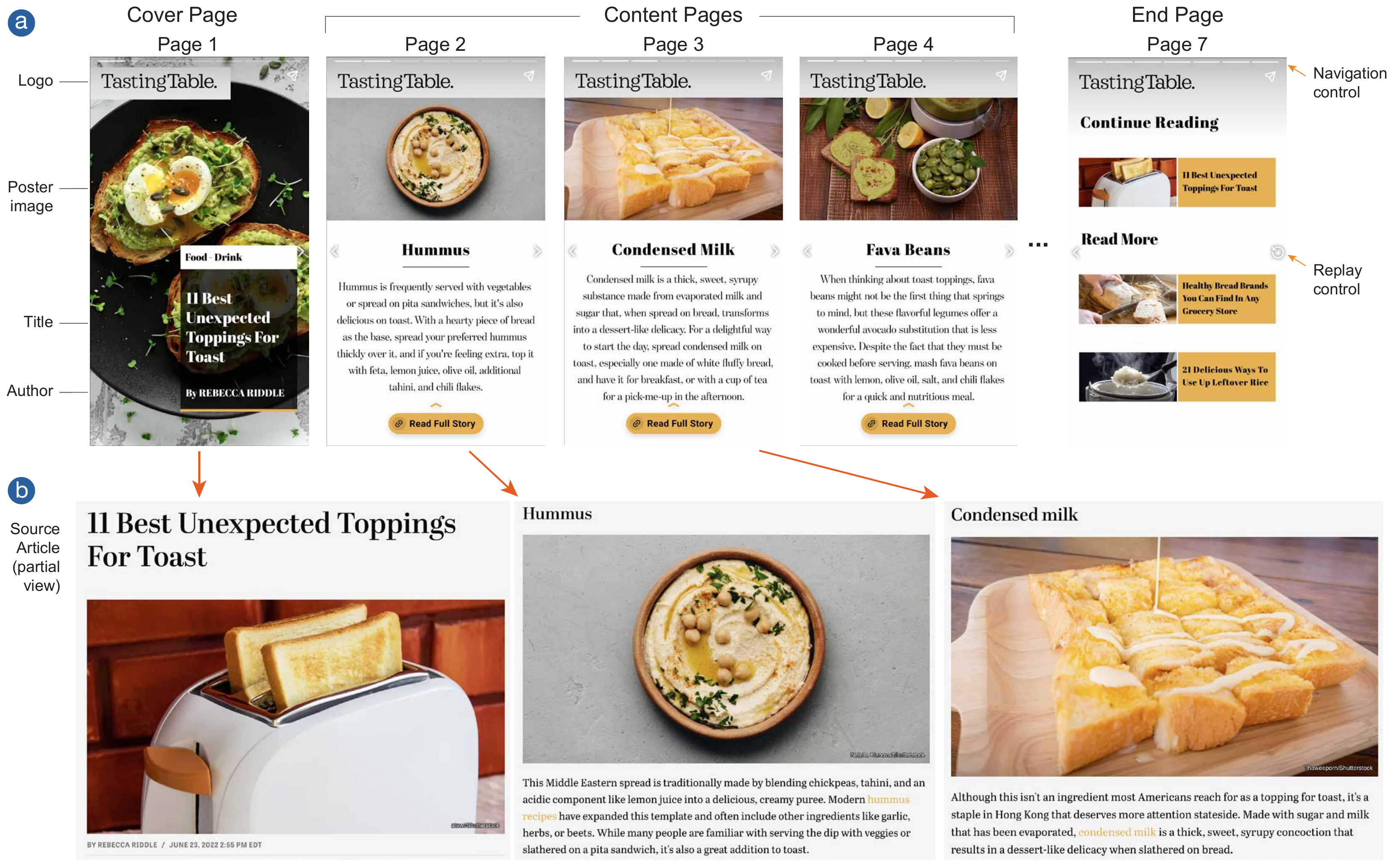}
  \caption{(a) A manually-curated Web Story, which starts from a cover page showing the title and author, followed by a sequence of content pages. It ends with a page of a list of article links. Users can navigate by tapping the navigational buttons or the progress bar. (b) The full article linked from this example Story.
  \emph{Source Story ``\href{https://www.tastingtable.com/stories/11-best-unexpected-toppings-for-toast/}{11 Best Unexpected Toppings For Toast}'' and \href{https://www.tastingtable.com/905522/best-unexpected-toppings-for-toast/}{its source article of the same title} by TastingTable, last retrieved in January 2023.}
  }
  \label{fig:example}
\end{figure*}

To further support wider audiences who prefer visual content~\cite{GoogleConsumerInsights}, recent studies suggested methods that convert structural articles to video presentations for engagement~\cite{TextVisualSlideshows2020,Videolization_Springer2018, HowToCut_UIST21}. \mobilehci{These methods visualize text content with relevant images for learning. However, learners often find the volume of content overwhelming to consume or prefer to follow content at their own pace with navigation controls~\cite{Chi:2012:MAG:2380116.2380130,SceneSkim_UIST15,MakeupBreakdown_CHI21}.}
%linear formats such as videos~\cite{Videolization_Springer2018, URL2Video_UIST20} force the viewer to watch at the pace of the video and are not adequate to provide an immersive, engaging, and mobile-first experience from long text~\cite{fiorella2018works, morrison2000animation}. 
% 
A recent trend in media consumption is to present structural content as a Web Story~\cite{navio2021guiding}, which contains a list of visual-driven web pages built on the AMP framework for viewers to tap through. It is especially suitable for mobile experiences~\cite{fiorella2018works, morrison2000animation} and is popular on modern platforms, including Facebook, Instagram, Snapchat, and Google~\cite{IG_story_doc19,SnapchatStories_chi17, WebStoriesAMP}. \mobilehci{Different from a slideshow or a short video, design guidelines suggest presenting quality images and concise text in 10-20 pages of a Web Story~\cite{WebStoriesPracticesAMP} as a digest, and readers can learn the full content from the source article (see Figure~\ref{fig:example})}. Creating a successful Story is a manual and challenging process. For a long article, creation involves designing and generating one or more Stories, along with organizing text and images in individual graphical layouts.

In this paper, we introduce \systemname, an automatic approach that converts a Wikipedia article of any length \mobilehci{and structure into a set of interactive Web Stories presented as digestible snippets, especially suitable for mobile viewing} (see Figure~\ref{fig:teaser}). 
% Our goal is to make long text content easy to follow and engage with for viewers seeking accessible, mobile-first, and visually-appealing presentation modalities. 
\systemname\ takes a structural document that contains multimedia assets, including text and images. It performs text summarization on each article section, leverages semantic similarity between article image features and the summarized output to select best matching visual assets. Finally, it creates multiple navigation paths based on the document hierarchy.
%for exploring the article by splitting a tree representation of the article hierarchy. 
% 
\mobilehci{A main Story introduces the article topic and links to a list of Section Stories. Each Story includes segmented text summary, images, and a voiceover presented in a suitable graphical layout. Viewers can navigate at their pace and follow the source article for in-depth content while \systemname{} maintains the mapping and structure.} 
% The Story provides options for the viewer to explore specific portions of the article that they care about. The Story also provides voiceover options for viewers who prefer listening to articles while observing other assets on the page as have been shown with instructional videos.

We evaluated \systemname{} by automatically generating 2,904 \added{multi-level} Stories from \numpagesrun{} Wikipedia articles of varying lengths covering four topic categories. We further explored select results with average viewers (N=14) and designers (N=3) via three user studies. \mobilehci{We compared user experiences between desktop and mobile Wiki2Story outputs with desktop and mobile application versions of the articles.} Users showed preference for our Stories in its ease of engagement, content understanding, and how pleasant it is to read compared to Wikipedia articles via the browser or mobile application. Our findings show that (a) our pipeline generates effective Web Stories from Wikipedia articles and (b) Web Stories are a promising mode of consuming long text content. Our work makes the following contributions:
% The quality of our output Stories are comparable to Web Stories manually created by designers using standard design tools.
\begin{itemize}[leftmargin=*]
\itemsep 0em 
\item A set of design principles for \mobilehci{automatic Web Story creation} drawing from our analysis of 50 manually-created examples.
\item An automatic approach to generate interactive Web Stories that dynamically summarize articles of varying lengths,
% \item Methods to convert hierarchical long text structures into a digestible format 
\mobilehci{where the content is comprehensible and consistent with the source article with quality visual presentations.}
\item An approach for image selection that leverages distance similarity metrics between article image features and summarized text for image matching while balancing image and article structure requirements.
\item Findings from user studies that validated the effectiveness and quality of our automatically-generated Web Stories.
\end{itemize}

\section{Related Work}
% \systemname{} builds on prior research on computational content creation, text summarization, and interactive multimedia navigation systems. We review these research areas and discuss their relationship to our work.

\subsection{Content Transformation}
Researchers have developed automatic approaches for creating interactive multimodal content from web articles~\cite{TextVisualSlideshows2020,URL2Video_UIST20, HowToCut_UIST21, doc2video_uist22} or text input~\cite{WriteAVideo_SIGGRAPH2019}.
% These systems provide new avenues to consume and understand content while addressing the challenge of creating such content at scale. 
Mackenzie \etal{} presented a system that transforms text articles of shorter length into audio-visual slideshows by leveraging word concreteness for image search~\cite{TextVisualSlideshows2020}. URL2Video uses constraint programming to convert design elements from a web page to a short video~\cite{URL2Video_UIST20}. HowToCut combines image and video assets with a voiceover from a structural tutorial into an edited video~\cite{HowToCut_UIST21}. We build on these works to develop an automatic approach that handles dense text content in a written article. \mobilehci{Commercial websites, web tutorials, or slideshows are commonly designed for fast consumption and have comparably sparse text with adequate images. In this work, we focus on articles that contain a significant amount of text sentences and hierarchies of varying depth. We convert such articles into a set of interactive Stories that preserve the article's meaning while considering visual quality and mobile browsing}. % while leveraging the hierarchical structure for navigation.

Our work is closely related to authoring tools for document summarization and presentation. 
% the Videolization system~\cite{Videolization_Springer2018} and Li \etal{}'s spoken dialog hierarchical summarization model~\cite{DialogSummarization_UIST21}. 
Wikipedia's VideoWiki project requires authors to summarize and script each scene to compose an output video~\cite{VideoWiki}. Videolization visualizes Wikipedia text content as a video using a knowledge graph~\cite{Videolization_Springer2018}. It is constrained to a few sample articles and performed extractive summarization of the Wikipedia document as a whole, which could lead to a significant loss of details across the article. In contrast, our approach generalizes to Wikipedia articles of any length and retains an article's meaning and structure by performing summarization at the section level. \mobilehci{Inspired by recent creativity tools, our work presents a first look into automatically converting web articles to interactive Web Stories with the goal of creating easy-to-digest content.}
% Li \etal{} focused on summarizing spoken dialog in a way that addresses challenges around speaker disfluency, lack of structure, and informality~\cite{DialogSummarization_UIST21}. Conversely, \systemname{} attempts to use the inherent structure of written prose to automatically create interactive Stories. Although we focus on Wikipedia articles, our approach can be applied to a wide range of structural long-text documents.

%\subsection{Text Summarization}
\subsection{\mobilehci{Text and} Multimodal Summarization}
Automatic text summarization entails producing a concise and fluent summary from a body of text while preserving key information~\cite{TextSummarizationSurvey2017}. Summarization methods have widely apply to web pages~\cite{WebSiteSummarization_ICIKM2000, WebDocSummarization_ACMHypertext2003}, scientific articles~\cite{SciLitSummarization_ACLHLT_2008}, emails~\cite{EmailSummarization_RANLP2003}, web search results~\cite{SearchSummarization_SIGIR2007}, and dialogues~\cite{DialogSummarization_UIST21}. The output of a text summarizer can be either \emph{extractive} (when the model returns informative sub-sentences from the input) or \emph{abstractive} (when the summarizer covers the main information fluently but may generate new words). 
% In the extractive case, the model returns informative sub-sentences from the input. 
% For abstractive summarization, the summarizer covers the main information fluently but may generate new words. 
% 
Early research explored single-document extractive summarization by computing a sentence score based on features such as the presence of keywords~\cite{edmundson1969new}, position in text~\cite{baxendale1958machine}, or word and phrase frequency~\cite{luhn1958automatic} and extracted sentences as summaries.

By applying machine learning~\cite{kupiec1995trainable, chuang2000extracting, mani1998machine} and natural language processing techniques~\cite{hovy1999automated, witbrock1999ultra}, recent works have leveraged Transformer-based sequence-to-sequence models on text generation and abstractive summarization tasks~\cite{dong2019unified, song2019mass, rothe2020leveraging}.
%for multiple Wikipedia articles ~\cite{liu2018generating}.
Liu et al. ~\cite{liu2018generating}'s approach automatically generates Wikipedia articles by summarizing multiple source documents.
% utilize a pre-trained model to perform summarization across multiple Wikipedia topics. We 
% \systemname{} adopts the state-of-the-art text summarization model, 
Zhang et al. introduced PEGASUS, which uses a pre-training self-supervised objective called gap-sentence generation to achieve human-quality summaries~\cite{zhang2020pegasus}. 
% By masking important sentences from an input document and generating missing sentences from the rest of the document, 
PEGASUS generates less abstractive text summary, which is especially suitable for Wikipedia content where retaining facts are important and less hallucination is required.
% PEGASUS performs well for summarization tasks. 
% video montage~\cite{WriteAVideo_SIGGRAPH2019}, text instructions as an interactive guide~\cite{HelpViz_UIST21}, web page to video~\cite{URL2Video_UIST20, HowToCut_UIST21},

\added{While there are research methods for creating multiple summaries from an article~\cite{modani2016generating} and presenting distinct summary perspectives~\cite{oh2008generating}, our work focuses on structure-sensitive summaries that are factual and preserve the correctness and relevance of the generated interactive Web Stories.}
\mobilehci{Multimodal summarization tasks generate textual summaries along with other media forms, typically images. Recent works have converted scientific documents to slides called Doc2PPT~\cite{fu2022doc2ppt}, generated textual summaries with images for news articles~\cite{zhu2018msmo, li2020vmsmo}, and visualized summaries of blog posts~\cite{bian2013multimedia} and complex sentences~\cite{uzzaman2011multimodal}. \systemname{} extends work in multimodal summarization research for structured articles where multimodal summaries are required to maintain a specific narrative hierarchy, and introduces these summaries to a new presentation form, Web Stories, for mobile browsing. This specific aspect makes our work unique from existing multimodal summarization research.}

\subsection{Interactive Navigation of Multimedia Content}
Previous works have demonstrated approaches for presenting content to viewers in a way that supports engagement and improves information retention. Interactive methods help viewers explore complex documentations via content highlight~\cite{DocWizards_UIST05} or extraction for sensemaking~\cite{fuse_uist22} \mobilehci{and accessible navigation~\cite{SlideGestalt_CHI23}}. To support instructional video content, a system can provide navigation path traces~\cite{kim2014data}, concept-map visualizations~\cite{tang2020supporting}, and time-aware word clouds for specific video segments~\cite{NonLinearNav_IUI2015}. Other systems have combined multimodal cues from text, audio, video~\cite{VisualNav_MM2017} and user activity~\cite{Replay_CHI2019} to direct users to the relevant part of a video or adapted content from existing videos for mobile experiences~\cite{Kim2022FitVid}. 

We focus on presenting dense informational content as a series of mobile-first Stories for viewers while allowing them follow specific narrative paths of the article. Our approach considers the amount of information in the entire Wikipedia article and the distribution of content within each section. The pipeline creates a single or multi-path Story based on the article length, and determines appropriate Story length according to the size of each section.

\section{Understanding Web Stories}

Web Stories are a publishing format introduced in the year 2018, designed for mobile devices to enable immersive, full-screen experiences across popular screen-based devices~\cite{WebStoriesAMP}. \mobilehci{Formerly known as the Accelerated Mobile Page (AMP),} it is an open-sourced development framework based on web technologies and has been adopted by popular publishers and content serving platforms\mobilehci{, including news, guides, tutorials, and personal stories. The framework defines the HTML tags for content (e.g., \code{\textless amp-story-page\textgreater} and \code{\textless amp-image\textgreater}) and structures (e.g., \code{\textless amp-story-grid-layer\textgreater}) with responsive styles, which enable authors to focus on constructing the content where the framework ensures consistent UX. According to product research in 2021, \userquote{Over 20 million Web Stories are already online, with 100,000 new Stories being added daily. And people on 6,500 new domains have published Web Stories since October 2020}~\cite{WebStoriesStats}.}

Driven by the need for fast content access and interactivity, a Web Story supports multimedia (including text, images, videos, audio, and animation) in a page-by-page presentation format. \mobilehci{Viewers can navigate by tapping on a touch screen or clicking via a point-based interface.} Researchers have highlighted how the Story format offers an avenue for sharing content through visual and full-screen storytelling while directing the narrative flow~\cite{navio2021guiding}\mobilehci{, which improves the quality of experience for mobile users~\cite{AMPUp_2019}. Unlike a slideshow, design guidelines suggest making Stories ``snackable'' with concise content. For example, a page should be restricted to 10 words and a video to take less than 15 seconds~\cite{WebStoriesPracticesAMP}. To concretize these constraints, we perform an analysis of published Web Stories and discuss the design principles for automatic Story creation.}

\subsection{Analysis of Web Stories}

To understand how a Story is different from slideshow presentations investigated by prior work~\cite{TextVisualSlideshows2020,TilingSlideshow2006,iSlideShow2010,fu2022doc2ppt}, we reviewed the principles for effective Stories creation defined by designers~\cite{WebStoriesPracticesAMP,WebStoriesStats}. Since we aim to convert existing articles to Web Stories, we further analyzed 55 published Stories and their corresponding source article. We randomly sampled Stories served by Google Discover on Android devices in July 2022. We selected Stories of any topic that had a corresponding article in English (i.e., with an exact or a loose one-to-one mapping of the title, sections, and content), linked from the Story. We annotated each Story's presentation, text, and links and compared the Story structure with its source article.
% Story formats are lightweight mobile-first interfaces that are increasingly becoming a major mode of information sharing. Navío‐Navarro and González‐Díez~\cite{navio2021guiding} highlighted how the Story format offers the possibility of sharing content through visual and full-screen storytelling while allowing the user to direct the flow of the narrative. Story formats contain several screens as part of an interactive timeline containing images, voiceover, videos, and short portions of text. For this work, we adopt web Stories as an appropriate complement to existing Wikipedia articles and an alternative to more complex video formats that are often difficult to load and much less flexible to dynamically control.
% 
Figure~\ref{fig:example} shows an example Story and its corresponding source article. Appendix ~\ref{appendix:existing_stories} contains more detailed analysis of the 55 Web Stories we reviewed.
The topics of these Stories covered cooking (40\%), gardening (21.8\%), travel (16.4\%), and others (technology, product reviews, sports, etc.; 21.8\%)

\subheading{Story Length.}
A Story consists of a series of pages, including a cover page, several content pages, and an end page. The design guidelines suggests, \userquote{An average length of 10 to 20 pages enables most authors to tell a good narrative}~\cite{WebStoriesPracticesAMP,WebStoriesStats}. From our analysis, a Story contains 8.6 pages on average ([min, max] = [5, 19]).
A navigation bar is present to indicate the length of a Story and the current page index. %(see Figure~\ref{fig:example}a top). 
\added{Ideally, a page contains less than 200 characters and can visually direct users to an external resource that contains the full information~\cite{WebStoriesPracticesAMP}.}

% \begin{table*}[t!]
% \caption{Design principles of Web Story creation from an article, derived from professional guidelines and our analysis of 55 Stories.}
% \centerline{\includegraphics[width=.4\textwidth]{figure/placeholder.png}}
% \label{tab:design_principles}
% \end{table*}

\subheading{Presentation.}
The cover page in each Story contains one quality image, with a title text overlaid on the top (54.5\%) or positioned above (27.3\%) or below (7.3\%) the image. 
In a Story, content pages commonly have the same font choices and layout type, which shows a few sentences and one to two images, sometimes with a page title. Consecutive pages may have slightly different layouts for engagement, such as swapping the text position from above to below the image. 
We observed that most Stories follow the guidelines~\cite{WebStoriesPracticesAMP} to present quality visuals, concise text sentences, and contrasting text colors for reading. 
% To address broad feedback on what makes AMPs successful, we will analyze a select number of AMP stories created by professionals. We will look at qualities such as their overall structure, use of multimedia, including images and text; and show how these inform our design decisions. 
% 
Most images are high resolution, with content mindfully cropped and expanded to fill the entire screen in the portrait view.
The end pages typically show either a link to the full article (61.8\%) or a list of more links to relevant articles (38.2\%).

\subheading{Mapping to the Source Article.}
We selected only Stories with a link to its source article and observed that Stories consistently embed the same linked button on each content page with a tag ``Learn more'' (50.9\%), ``Read full story'' (16.4\%), ``Read more'' (5.5\%), or topic-related actions (``Get the recipe'' or ``Let's Bake!'' for cooking).
The structure and ordering of a Story follows the source article. For example, each Story page for a recipe, a travel guide, or gardening shows a cooking step, a tourist attraction, or a gardening tip respectively.
Text content is often summarized from a paragraph, or simply preserves the first sentences.
Finally, most Stories directly reuse the same images from the articles, while some may apply web footage or repeat the same images.

\subsection{Design Principles}
From the Web Stories creation guidelines and our analysis, we derived four design principles for text-image article conversion to a Web Story. %(see Table~\ref{tab:design_principles}): 
% This presentation guides users to focus on the visuals for processing concise text information. 
\begin{itemize}[leftmargin=*]
\itemsep 0em 
\item \added{\textbf{D1: Succinctness and Coherence.}} A Story should tell a coherent narrative within reasonable length, between 4 and 10 pages. It should include \mobilehci{a structure for storytelling, starting from} (1) a \emph{cover page} that shows the title and a captivating image, (2) a list of \emph{content pages}, and (3) an \emph{end page} with one or more links to other Stories \mobilehci{for topic exploration}.
\item \added{\textbf{D2: Harmonious Flow between Pages.}} Each page should capture one idea with a supporting image, such as \mobilehci{to describe a step or a tip}. Consecutive pages should be harmonious without a sudden transition in topics.
\item \added{\textbf{D3: Consistent Visual Layout Presentation.}} The presentation in a Story should be consistent, with the same design choices (layouts, fonts) and images with similar quality.
\item \added{\textbf{D4: Quality Images and \mobilehci{Readability}.}} Images should be high quality and if needed, cropped properly \mobilehci{to focus on the subject}. Text should be concise with a reasonable font size for reading, limited to one or a few sentences per page.
\end{itemize}

%\subheading{Creation Method.}
To our knowledge, creating a Web Story is currently a manual process via GUI editors and mainstream design tools~\cite{WebStoriesEditor}. There are existing plugins to pull content from web articles for human editing. However, it requires multiple iterations to select text and image materials, summarize paragraphs, match with relevant and quality images, design and populate the layout.  \mobilehci{Finally, it may take additional edits to review the final rendered Story and make adjustments (e.g., anchor to a specific page and adjust its image.)} This can be a time-consuming process \added{for content authors, where each of the aforementioned steps could take multiple work hours over several days, especially to create a multi-page Story from a long article}.
Therefore, we propose a method for automating the creation process for text documents leveraging computational analysis and design techniques.
% When needed, offering additional context via ``Learn more''

\begin{figure*}[th!]
  \centering
  \includegraphics[width=.9\linewidth]{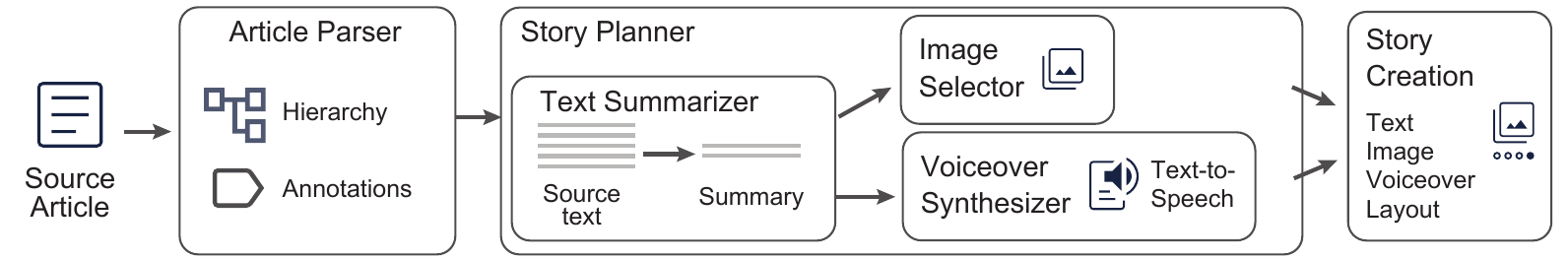}
  \caption{\systemname{}'s pipeline consists of an Article Parser that retrieves the input article hierarchy, a Story Planner that summarizes sections and designs multiple Stories, and Story Creation to present images and text in graphical layouts.}
  \Description{\systemname{} ...}
  \label{fig:pipeline}
\end{figure*}

\section{Mobile Browsing with \systemname{}}
We present \systemname{}, an automatic approach that converts a Wikipedia article of any length into a set of interactive Web Stories suitable for mobile consumption. Our end-to-end pipeline parses the hierarchy of an input article to identify sections, each generating a text summary. It plans multiple Story paths, selects image assets that best support the content, \mobilehci{and composes multimedia content} in visually appealing layouts. 
It generates a \emph{Main Story}, which consists of a series of pages that illustrate the overview of the article. 
%Each page is a single interface within a Story. 
The end of the Main Story points to a list of \emph{Section Stories}, each leading to a unique narrative path of the input Wikipedia article.

To illustrate \systemname{}’s viewing experiences, assume that a user, Onyeka, is interested in learning \mobilehci{a topic about apple fruit} from a Wikipedia article titled \textit{\mainexample{}}, for her science project (see the example in Figure~\ref{fig:teaser}). 
% \footnote{Wikipedia, \href{\mainexampleurl{}}{\mainexample{}}, \url{\mainexampleurl{}}} 
% Onyeka identifies, either directly or through a web search, a Wikipedia link that contains information about the topic. She then opts to explore the topic in a Story format created with \systemname{}. 
On her mobile device, Onyeka sees that the Main Story presents the article’s overview section as six pages, each with a text snippet of one to two sentences overlaid or placed next to an image.
The first content page shows a quality image of a complete apple. The text below shows, \userquote{An apple is an edible fruit produced by an apple tree (Malus domestica).} Onyeka reads through the pages, where each page has an optional Text-to-Speech voiceover feature to guide her at a steady pace.
Compared with reading the source article, she could digest the content by reading the paragraph breakdowns with visual and auditory support. This helps her better follow the terms or jargons that she may not be familiar with, including the mention of \userquote{species} (in Page 3) and \userquote{ancestor} (in Page 4).
% with visual. The output presents an initial Story that contains three to five content cards from the article's description section. A content card is a type of card that contains a text snippet and an image from the Wikipedia article with accompanying voiceover; the first and last cards in the output stories are not content cards

At the end of the Main Story, Onyeka sees a list of options, each linked to an article section, including \emph{History} and \emph{Uses}.
\mobilehci{Building upon the Web Stories open source framework, she finds consistent interaction to navigate between pages and follow links on her mobile phone.} Onyeka taps a topic to continue learning more from the article, where each Story ends with the options to navigate to other sections for further exploration.
\systemname{} tracks the progress as she reads and provides a visual hint.
She could also visit the corresponding section in the source article by following the ``Read Full Article'' link on the Story page.
% Onyeka has control over the order in which she explores the sections or if they even wish to only see a specific section.
% To enable a better viewing experience, we focus on Wikipedia articles that contain at-least five images within its sections. 

\section{Story Creation Pipeline}

\systemname{} consists of three major components to automatically convert a Wikipedia article to a Web Story (see Figure~\ref{fig:pipeline}): (1) An \emph{Article Parser} that retrieves and parses a Wikipedia document, (2) a \emph{Story Planner} that organizes text, images, and page information. It uses a text summarization model and selects images to generate layouts for the final (3) \emph{Story Creation}. We describe the detailed design of each component.
% First, it retrieves and parses a Wikipedia document containing information about each element in a given article. It selects the texts, headings, images, and page information. For text content, \systemname{} passes the text for each section through a text summarization model and adds narration for each summary snippet. It then assigns images to text based on existing associations with the article.
% \systemname{} analyzes the hierarchical structure of the article to build a tree structure and decides on a Story planning strategy based on the article length (and tree depth). \systemname{}'s presents the information using a visual layout system that creates designs for each page. \systemname{} then uses the generated Story plan and the created layout designs to create AMP story formats of the Wikipedia article.

\subsection{Article Retrieval and Annotation}
We developed a data extraction pipeline that retrieves metadata from MediaWiki-based inputs~\cite{MediaWiki} to access Wikipedia data. \mobilehci{\systemname{} retrieves the page information (page title, page description, and the language), links, and complete sections of each article. 
% From this data storage, we access three categories of data about an article: first, general page information such as the page title, page description, and the language in which the article is written. 
% 
Next, it parses each section at all levels for the corresponding title, section text, and index according to the article hierarchy. 
%Every section has a parent index except the description section of the article which is the root section. Every level 1 section in the article has the root section as its parent, and every section is a parent of any subsections within it. We assign a boolean value on whether the section is a content section or not. Wikipedia content sections are article sections that contain text directly related to the article and may or may not contain accompanying media.
It filters non-major sections such as ``See also'', ``References'', and ``External links'' to avoid generating non-critical Stories.
%that are not part of the article topic.
% 
To respect the copyright, \systemname{} only selects image assets that are under Wikipedia's shared license. It records the metadata of each image, including its resolution, the index of the contained section, and the source link.} %with the corresponding section.

% ------------------------------------------- %

\begin{figure*}[t]
  \centering
  \includegraphics[width=0.8\linewidth]{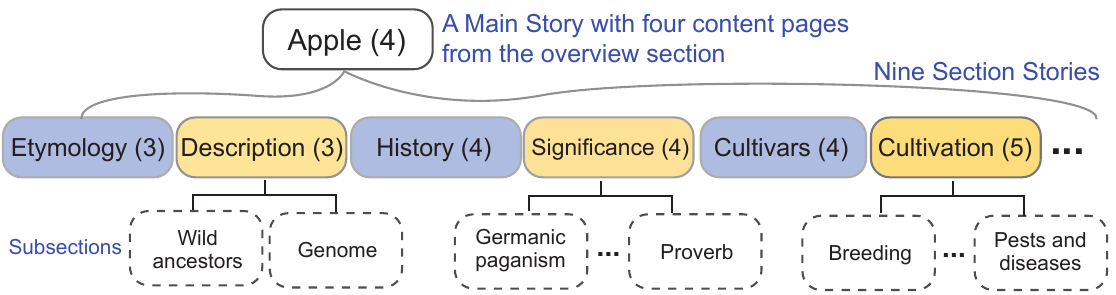}
  \caption{\mobilehci{Considering the mobile browsing experience, }\systemname{} dynamically generates a set of Web Stories based on the document hierarchy \mobilehci{to ensure each Story is coherent and easy to consume}. Each section generates an individual Story. For sections with no subsections (shown in color blue), \systemname{} uses the SPLIT strategy to break its content into multiple pages. If a section has one or more subsections (shown in color yellow), it uses the MERGE strategy to incorporate all its children section as individual pages in the Story. The section number shown in this tree denotes the number of content pages in a section Story.}
  \Description{\systemname{} ...}
  \label{fig:paths}
\end{figure*}
% It uses two methods, SPLIT and MERGE, to balance the Story length given the section structure.

\subsection{Text Selection and Summarization}
In order to provide easily digestible content while ensuring reasonable coverage of the article, we perform text summarization at the section level. 
We adopt the text summarization language model called PEGASUS~\cite{zhang2020pegasus}. 
% PEGASUS has shown notable improvement over existing summarization methods in terms of the quality of its output summaries. 
% Its model is trained on multiple domains, including news articles, Quora, and Reddit forums. 
% However, some of these PEGASUS variants either inaccurately over-abstract~\cite{FaithfulSum2020}, hallucinate~\cite{DialogSummarization_UIST21}, or return only one sentence summary outputs. 
% 
Given the context of Wikipedia, we chose to use a variant of the model that was trained on CNN and Dailymail dataset %(\textit{cnn\_dailymail}) 
% , which contains 93k articles from CNN and 220k articles from Daily Mail
~\cite{NIPS2015_afdec700}, \added{which has a higher similarity to Wikipedia articles, compared to other models trained on social media content such as Reddit posts, for example.}
% Though the model produces longer summaries and did not hallucinate, it tends to do fewer abstractions and returns summary sentences that might look like some in the original input.
One advantage of this model is that it better maintains the original input with fewer abstractions, which avoids too many spurious additions for articles containing factual information.
% 
% We iterate over each content section in an article and pass its constituent text as the input to the model. 
 
\mobilehci{Wikipedia guidelines suggest that contributors use strong opening sentences in a section~\cite{WikipediaFirstSentence}. We maintain this strength by passing the input text minus the opening sentence to the summarizer. We then concatenate the first sentence to the summarizer output to make our final summary output. We did not summarize sections with less than 50 words. If a section contains multiple subsections, we create summaries for the text in each subsection, up to the lowest level. We converted the full text content of such sections to a Story.}
% 
%In implementation, we made design decisions to (1) maintain the first sentence unchanged, followed by the summary, and (2) maintain a short section that is less than 50 words.

% ------------------------------------------- %

\subsection{Story Planning}
Based on the article hierarchy and text summary, \systemname{} organizes the content into one to multiple Story paths. 
At a high level, we define two types of sections: the \emph{overview section} as the root of the article structure, which we convert to the \textbf{Main Story}, and \emph{the content sections} form the first level of the tree, which we convert to narrative paths as \textbf{Section Stories} (see Figure~\ref{fig:paths}). % and serve as both the children of the root node and the parents of their constituent subsections. 
% To maintain consistency with the Wikipedia article, we adopt these level 1 section nodes as the headings for each narrative path of the article. 

Drawing from \mobilehci{the design guideline of the story length from Section 3.2 (D1), we define two types of creation strategies} depending on the length of the original article. Denote the number of content sections of an article as $s$ and the maximum number of pages allowed in a Story as $n$. We adopt an adaptive approach to create either a ``Compact Story'' that has no navigation path when $s \le n - 2$, or a ``Multi-Path Story'' for longer articles. The two constant pages are reserved for the cover and the end pages.

% , based on the guidelines~\cite{WebStoriesPracticesAMP}.
% Considering an input article's length greatly varies, w

% we analyze the number of content pages that can be created based on the number of content sections in the article. 
% For an article of $s \le n - 2$, where 

% , we assign a Compact Story format. Otherwise, we assign a Multi-Path Story format to the article.
% for scalability:
% \subheading{Handling various lengths.} 
% There is a trade-off between maintaining the consistency of the Story with the Wikipedia article's structure and finding an appropriate number of pages for the Story. Metrics such as the overall length of the article, the number of sections, the number of subsections per section all vary. To address this trade-off, we make two design decisions. 
% First, we investigate when a story should be a ``Compact Story'' with no additional navigation paths. 
% 
% Second, we address differences in length between navigation paths where a section has significantly more depth than another section within the same article.
% For design decision 1, 
% \subheading{Compact Story:}  
% 
To create a Compact Story, we perform a pre-order traversal of the article structure starting at the main section node and present each subsection as single page in the Story.
% For edge case where the number of content sections are less than \textit{n - 2} but some sections have considerable depth (greater than level 2 of the tree), we also assign a multi-path story format to the article. 
% \subheading{Multi-Path Story:} 
% To create a Compact Story, we perform a pre-order traversal of the article structure starting at the parent section node to determine the constituent pages within each Section Story. We allocate each subsection with a page. 
% For a Compact Story, we present each subsection as a page, sorted by the tree order.
% 
% In this way, each section makes up a Story that can be reviewed independently (see Figure~\ref{fig:paths}).
% For design decision 2, 
% 
For a Multi-Path Story, each section generates a constituent Story that presents the subsections within that section. A pre-order traversal of the article structure ensures that all subsequent lower-level children are captured as individual pages in the Story.
We set the maximum number of pages in a story, $n=10$, based on our analysis and recommended best practices~\cite{WebStoriesPracticesAMP,WebStoriesStats}.

The section length of a Story varies depending on the number of subsections. \added{If a section has one subsection, we can split its content to a multi-page Story. If a section contains multiple subsections, we can constraint each subsection to one page. To this end}, we propose two page creation methods for a Multi-Path Story in order to balance the page length in line with our principles:
(1) \emph{MERGE}: When an article section has subsections, each page in the Story represents a summary from each of the constituent subsections. We apply these subsections as pages in the section story. A page may therefore contain multiple sentences from the summary (see the final section \userquote{Uses} in Figure~\ref{fig:teaser} that has eight subsections for example.)
(2) \emph{SPLIT}: When an article section has no or one subsection, we split the section summary text to fit multiple pages (see the section \userquote{History} in Figure~\ref{fig:teaser} for example). This keeps section stories within a minimum number of pages for a better navigation experience.
% If this condition is true, we adopt a SPLIT strategy where each summary sentence makes a page.

% Take Figure~\ref{fig:teaser} for example. The final section \userquote{Uses} has eight subsections, from \userquote{Popular uses} to \userquote{Toxicity of seeds}, each as a page.

% ------------------------------------------- %

\begin{figure}[t]
  \centering
  \includegraphics[width=0.7\linewidth]{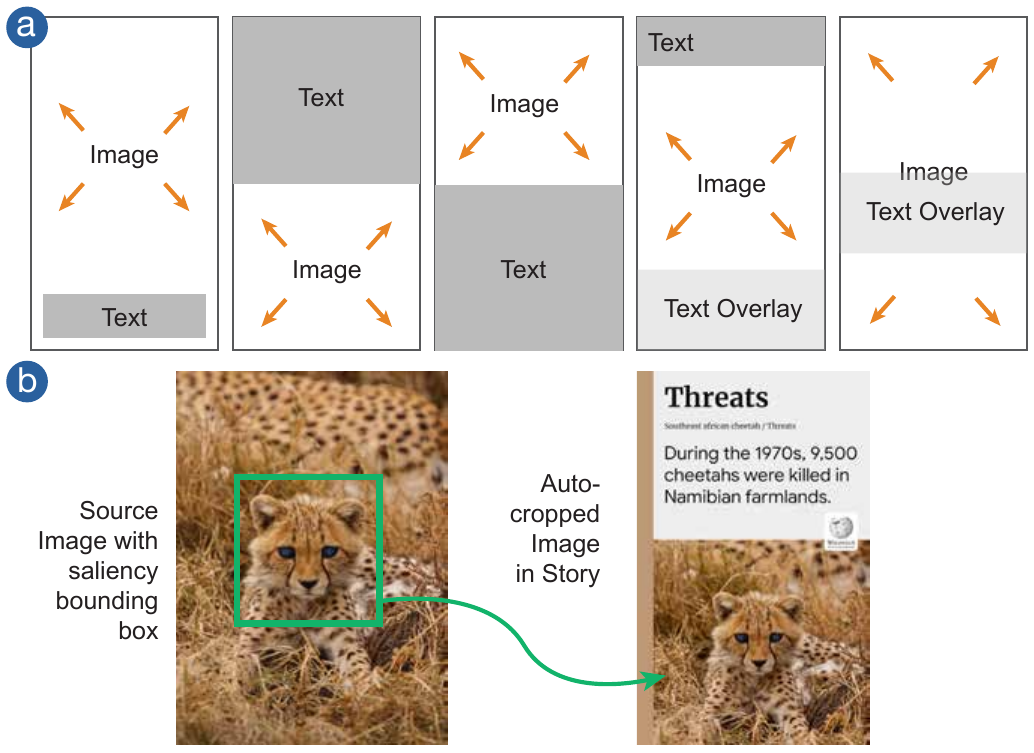}
  \caption{\systemname{} automatically selects suitable layouts for text and visual assets by considering the image quality and text length (a). It positions an image to focus on the most salient region (b). \emph{Image source: ``\href{https://en.wikipedia.org/wiki/File:Baby_Cheetah_-_Marko_Dimitrijevic.jpg}{A stunning baby cheetah}'' by Marko Dimitrijevic is licensed under CC-By SA 2.0.}}
  \Description{\systemname{} ...}
  \label{fig:layout_generation}
\end{figure}

\subsection{Content Presentation}
Each of our generated Story contains three types of pages: (1) a \emph{cover page} that presents the article title, (2) a \emph{content page} that contains a heading, summary text from the article, a supporting image, and a link to the source article, and (3) an \emph{end page} for navigation. 
% A cover page introduces the topic and contains the article title with the article's leading image. 
% A content page contains a heading, summary text from the article, an accompanying image, and a path name specifying its hierarchy within the article. 
% A navigation page contains links to explore every other section of the article except the one currently active. 
Below we outline \mobilehci{our method of Story page generation}.

\subsubsection{Image Assignment}

\mobilehci{Our image assignment approach is similar to the image-text alignment technique presented by Chowdhury et al.~\cite{nag2021sandi} and Chu et al.~\cite{chu2017blog}. We select and align images from an image collection with text paragraphs by leveraging extracted deep image features and visual tags. To produce Stories that are consistent and semantically meaningful with the source article, our implementation respects the article hierarchy where the image appears in an article section.}

To select appropriate images for a Story, we utilize images from the Wikipedia Image-Text (WIT) dataset~\cite{Krishna_SIGIR2021}, which is a human-evaluated dataset of Wikipedia images and texts for training multimodal multilingual systems. 
The images from the WIT dataset are research-permissive licensed and have been filtered for quality and appropriateness.
% including removing images with height or width less than 100 pixels for enhanced quality, and retaining only images with research-permissive licenses. 
For a given article, \mobilehci{we maintain a list of images and the corresponding section where they are positioned. We pre-compute visual features for each image.} 

For a given story page, we match images to their corresponding section summary. If a section has no image in the original list, we assign an image from sections with more than one image or re-assign an image from an existing section. To select the best matching image, we compute an embedding distance between the section summary and the image features and select the highest matching image. To avoid repetition, if the highest matching image is the most recent from a preceeding page, we select the next best image. \mobilehci{To ensure the fidelity of a Story with the original article, we did not include images outside the source article, while prior work has suggested generative or search approaches that we chose not to experiment~\cite{ImagenEditor_CVPR23,VisualCaptions_CHI23}.}

\added{Note that we experimented an alternative approach by pairing the sections with the available images and assigning dynamically based on overall match instead of sequentially. However, this did not produce a noticeable difference in results and was not adopted. We posit two reasons: 1) Most articles had fewer matchable images than sections, so that the most relevant images were likely to appear more than once regardless of the approach, and 2) Sections were more likely to be matched with images from a neighboring section, making a first best matching approach useful.}

%In a given story page, we matched images to their corresponding section in the . To increase the diversity of images, as . If a section has no corresponding image in image, we compute the 

%We maintain a sorted list of WIT images, each mapped to a section. We allocate new images to pages to increase image diversity (i.e., avoid showing the same images in Stories). If an article has few assets, we compute the presentation distance between images and re-allocate them to pages.
% For a target section, we assign the first image of the WIT images from the section, sorted by presentation orders. 
% 
% To increase image diversity (i.e., avoid showing the same images in Stories), we maintain a set of unused images for 
% When a section has multiple images, we use the extra images to supplement sections that have no image. If there are no extras, we rotate the given images among the pages.

\subsubsection{Layout Selection}
To enhance the visual variety of Stories, we build on recent works in graphic layout generation given design constraints~\cite{lee2020neural, arroyo2021variational}. We created two classes of layout templates for content pages based on the text length, including a \emph{short-text} class of eight designs and a \emph{long-text} class of six designs.
We compute the average length of all generated summaries in the input article and assign a template to a summary from either class. 
% If the summary length was less than the average length, we assign a design from the short-text class. Otherwise, assign one from the long-text class. 
The cover page uses the same layout for consistency (see Figure~\ref{fig:teaser} for examples.)

Each design template has an image, text, and decoration rule (see Figure~\ref{fig:layout_generation}a):
(1) The \textbf{image rule} specifies the best crop to fit a page using the deep neural network proposed in Creatism~\cite{fang2017creatism} (see Figure~\ref{fig:layout_generation}b). This centers the most relevant objects in the image.
(2) The \textbf{text rule} ensures that text fits properly within the page with adjusted font sizes with respect to the text length. 
(3) The \textbf{decoration rule} transfers the dominant color from the input image to the solid shapes within the template for visual harmony.
% It selects a major color from the image and applies to the decoration elements for visual harmony.

\subsubsection{Web Story Creation with a Voiceover}
% Given the required information for a Story: sentence summaries, assigned image, layout template, and story plan; 
We dynamically compose the text, image, and layout components using \mobilehci{the HTML-based Web Story framework for interactive, cross-device viewing}~\cite{WebStoriesAMP}. 
% for the main Story and all constituent section Stories.
%
The Stories are optionally enhanced with a Text-to-Speech (TTS) voiceover to help users auditorily follow the text summary. \mobilehci{We included this optional feature to the tool drawing on positive user experiences in prior related work for content following~\cite{URL2Video_UIST20, TextVisualSlideshows2020}, although we do not claim novelty on this aspect}. We synthesized each text sentence using Google's Text-to-Speech API~\cite{SpeechAPI} into audio MP3 files for interactive playback. 
% Since each summary paragraph is mapped to a content page, we check if the page employs a SPLIT or MERGE strategy. If a SPLIT strategy, we first split the summary paragraph into sentences and create voiceovers for each summary sentence. If a MERGE strategy, we create a voiceover for the summary paragraph as is. Though the default viewing length of a page is defined by the length of the voiceover, a viewer has the option of navigating in and out of pages or toggling off the voiceover should they wish.

% ---> Peggy's note: A lot of the descriptions are very detailed for implementation but not research focused, so I removed and tried to rephrase and make it concise.

% ------------------------------------------- %

% \subsection{Implementation}
% We implemented our end-to-end pipeline in C\texttt{++} and Python. %The text summarization model is a pre-trained version of the PEGASUS abstractive text summarizer~\cite{zhang2020pegasus}. 
% % The story planning and Wikipedia tree structure creation and traversal modules use the Treelib library.
% We used the Web Story framework~\cite{WebStoriesAMP} to generate HTML pages that can be accessible by conventional web browsers on a mobile device with adaptive experiences to screen resolutions.

\begin{figure}[t]
  \centering
  \includegraphics[width=0.7\linewidth]{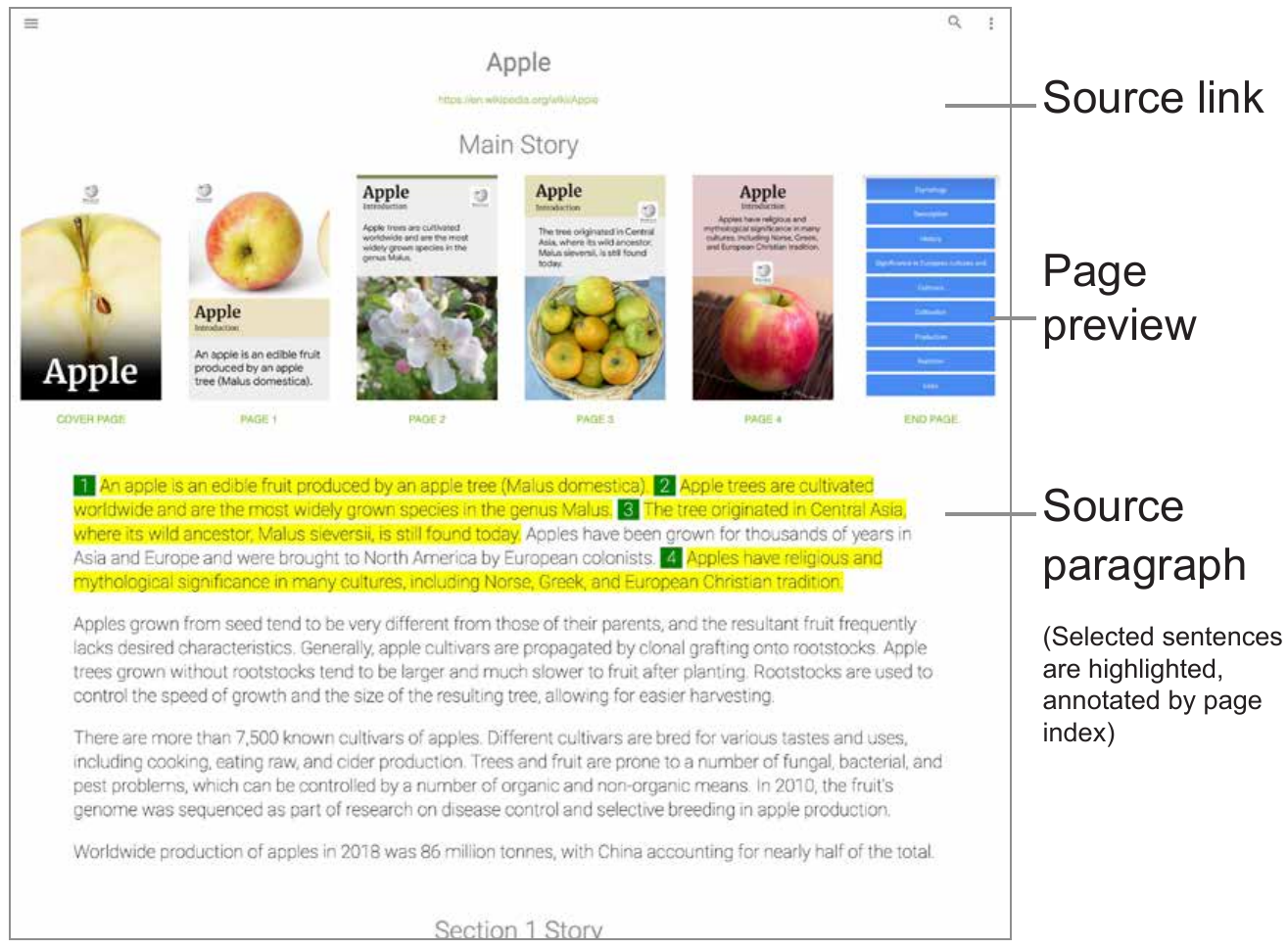}
  \caption{In our Review UI, inspectors can review the automatically-generated Web Story results, including the source text, images, and test layouts for each page corresponding to the source article.}
  \Description{\systemname{} ...}
  \label{fig:review_ui}
\end{figure}
% \begin{table*}[t!]
%   \Description[]{}
%   \caption{Analysis of \systemname{}'s input and output composition for \numpagesrun{} Wikipedia articles. Each row represents the average value of \numpagessampled{} articles for a specific category.}
%   \label{tab:wiki_analysis}
%   \begin{tabular}{lrrrrrr}
%     \toprule
%     \multirow{2}{*}[-4pt]{\thead{Category}} &
%     \multicolumn{3}{c}{\thead{Source Wikipedia Articles}} &
%     \multicolumn{3}{c}{\thead{Output Web Stories}} & 
%     \cmidrule(lr){2-4}
%     \cmidrule(lr){5-7}
%     & Sections & Words per section & Images & Summary words per section  & Main + Section Stories & Words per story page\\
%     \midrule
%     Animal & 22 & 211.1 & 14.9 & 58 & 6 & 15\\
%     Architecture & 14.9 & 155.5 & 7.4 & 52 & 5 & 16 \\
%     City & 18.5 & 185.5 & 12.7 & 54 & 5 & 16 \\
%     Plant & 23.9 & 172.9 & 10.4 & 52 & 5 & 16 \\
%     \bottomrule
%   \end{tabular}
% \end{table*}

\begin{table*}[t!]
\caption{Analysis of \systemname{}'s input and output composition for \numpagesrun{} Wikipedia articles.}
\centerline{\includegraphics[trim=50 380 50 65,clip,width=.9\textwidth]{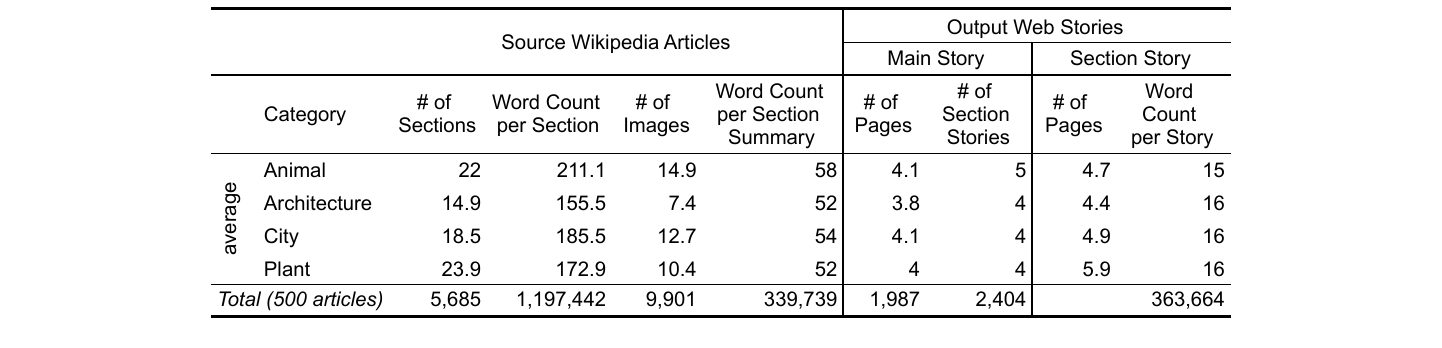}}
\label{tab:wiki_analysis}
\end{table*}

\begin{figure*}[t]
  \centering
  \includegraphics[width=\textwidth]{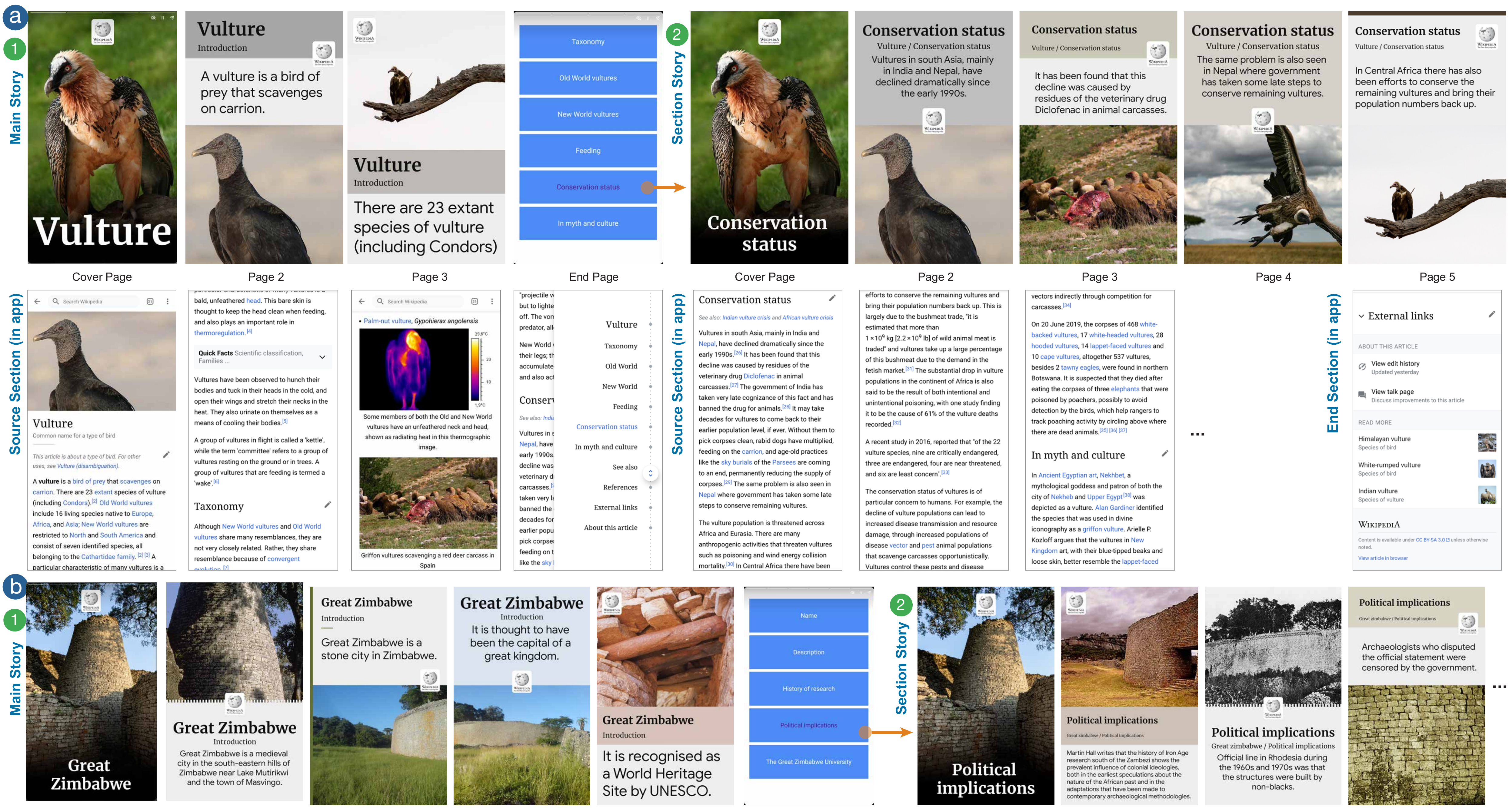}
  \caption{\mobilehci{Example results from \systemname{}'s automatic pipeline, compared to its source article presented in the Wikipedia app. \systemname{} summarizes paragraphs in a section and identifies relevant images in dynamic layouts for presentation. Each Story has a different length that is visible via Web Story's progress bar. We encourage the readers to follow links to the source for comparison.}
  \emph{Source articles: (a) ``\href{https://en.wikipedia.org/wiki/Vulture}{Vulture}'' and (b) ``\href{https://en.wikipedia.org/wiki/Great_Zimbabwe}{Great Zimbabwe}'' by Wikipedia are licensed under CC-By SA 3.0.}
  }
  \label{fig:results}
\end{figure*}

\section{Results}
To demonstrate the effectiveness of our pipeline and examine the quality of the generated Stories, we created a dataset of \numpagesrun{} Wikipedia articles and discuss the method and results below.

\subheading{Data and Method.}
We selected four high-level categories of popular topics for education: \textit{Animal, Architecture, City}, and \emph{Plant}. We focused articles with at least five sections and five images and these categories returned the most articles. We did a random sample from the Wikipedia dataset to select an equal number of articles from each category that fulfilled the criteria, making a total of \numpagesrun{} articles. \added{We selected \numpagesrun{} articles to test \systemname{}'s ability to generate Stories at scale. On average, these articles had 20 sections across multiple levels.}
%that 
% 
% \subheading{Method.}
%we ran the \systemname{} end-to-end pipeline. %for all articles.

% ------------------------------------------- %

\def\totalMainStories{500}
\def\totalSectionStories{2,404}
\def\totalMainPage{1,987}
\def\totalImages{9,901}
\def\totalSections{5,685}

\subheading{Generation Results.}
We were able to generate Web Stories for all articles (which included \totalSections{} sections and \totalImages{} images), containing a total of \totalMainStories{} Main Stories (which consisted of \totalMainPage{} pages) and \totalSectionStories{} Section Stories.
Table~\ref{tab:wiki_analysis} shows an analysis of the source articles and our generated outputs. \added{It took between 2 to 5 minutes per article to run through our end-to-end pipeline -- from performing the retrieval, summarization, planning, to layout selection. Among these components, the section-by-section text summarization required the most creation time, accounting for more than 70\% of time spent on story generation.} For each article, \systemname{} generated one Main Story (consisting an average of 3.97 pages) and an average of four Section Stories.

\subheading{Quality Observations.}
Figure~\ref{fig:results} shows example outputs from \systemname{}. 
\added{We sampled eight generated Stories in order to validate their adherence to the principles outlined in Section 3.2 (including \principle{D1}: Succinctness and Coherence, \principle{D2}: Harmonious Flow between Pages, \principle{D3}: Consistent Visual Layout Presentation, and \principle{D4}: Quality Images and Typography). The goal of this analysis was to provide supporting evidence whether the generated Stories are comparable to the manually-created ones that we had studied.}
To help inspect the automatically-generated Story results, we built a web-based Review UI that shows the source text and images from the input article (see Figure~\ref{fig:review_ui}).

\added{For \principle{D1}, we observed that \systemname{} effectively summarized the overview section of an article and created a Main Story with diverse images from the article (see Figure~\ref{fig:results}a-1 and b-1). The number of pages generated for the Main story is proportional to the length of the overview section. The structure of the generated Stories is consistent across articles and levels, with each Story containing a cover page, a set of content pages, and an end page linking to other stories.}

\added{For \principle{D2}, we noted that the language model's extractive summary of text snippets at the sentence level helped make a Story comprehensible and allowed a smooth transition between pages. For articles with layered sections, \systemname{} generated multiple Section Stories to allow navigation to specific parts of the article (see Figure~\ref{fig:results}a-2 and b-2).} 

\added{For \principle{D3}, we found the layout selection component of \systemname{} effective. The border colors of the layout matched the dominant colors of an input image (as shown in Figure~\ref{fig:layout_generation}b). \systemname{} selected multiple layouts to make the pages visually distinct, which made the content engaging. This was particular useful for structuring pages with fewer text assets (less than 10 words).}

\added{Finally, we inspected the image and text quality in a Story (\principle{D4}). Though the quality of the images have been pre-vetted in the WIT dataset, \systemname{}'s image cropping feature ensured that images were prominently displayed. The images were centered to the main subject properly. The layout balanced both the text and image at the same time, which is different from reading a text article. }

\subheading{Technical Opportunities.}
We identified two categories of cases for improvements in the generated examples. The first case involves Wikipedia articles with advanced data structures in the articles than text and images. Our pipeline excluded complex tables, multi-level lists, or structures such as cladograms. The second case was for articles with images that were not necessarily relevant to the section topic. This is due to the fact that a best feature match may not necessarily imply a perfect match for the summary in context. However, our pipeline was able to generate reasonable outputs from these cases. In general, we find that the automatically-generated Web Stories were consistent with our goal of creating concise and visually-appealing Story complements of Wikipedia articles.
\section{User Evaluation}
\added{Building on our validation of the generated results, we conducted three study sessions to evaluate the effectiveness and quality of our Web Story generation output. Specifically, we aimed to (1) understand how well \systemname{} presented useful information from a Wikipedia article to the reader; this is characterized by the ease of engaging with the output Story and enjoying the reading experience, and (2) evaluate the overall quality of the automatically-generated multi-path Web Stories}. We conducted three studies with average viewers and professional designers.
% by exploring how easy a user can locate relevant information via navigating the Story levels and understanding the summaries.}
%To gain insights on the \added{quality and} usability of our automatically generated multi-path Web Stories, we conducted informal evaluation sessions with three professional designers and eight additional participants.
%To gain insights on the usability of our automatically generated multi-path Web Stories, we conducted informal evaluation sessions with eight participants. In addition, we invited three professional designers for their insight on our results.

\subsection{Study I: Viewer Inspection \mobilehci{on Content Transformation}}
In Study I, our goal was to understand whether average viewers agreed that our Stories (as the \systemname{} condition) facilitated similar understanding as the source articles (as the Baseline condition). 
% 
% provided the baseline condition (that reviewed the Wikipedia web articles) and the Story condition (that reviewed \systemname{}'s generated output.)
% 
% We split the participants in two groups. The first group (Wiki-first) viewed a topic as a Wikipedia web article before being presented with an auto-generated web Story. The second group (Story-first) navigated the auto-generated web Story before being asked to view the topic as a Wikipedia article. 
\subsubsection{Study Design}
To inspect the content quality in a 30-minute session, we selected two automatically-generated Stories we validated in Section 6 where the source articles had a medium length. % and their corresponding auto-generated multi-path Web Stories. 
% The Stories contained a main story and multiple navigation paths. We removed the voiceover from the Stories but focused on the visual presentation. \added{This allowed the participants evaluate the visual modalities without the added input from the text-to-speech.}
% 
We counterbalanced the order and asked participants to review the content as if they were researching the topic rather than casual reading.
% interact with each format in the pre-assigned order. % The tasks and questions were unchanged regardless of the ordering. 
In each condition, we asked participants to verbally describe their strategy for learning the topic. We also verbally prompted questions and asked them to locate the information that could be found in a non-overview section or in a Section Story.
% The questions were curated such that their answer is found in both formats.
\mobilehci{Since the goal was to verify content transformation of lengthy source articles, we chose to facilitate the remote study where participants reviewed on their desktop or laptop via keyboard or mouse control, where we leave the mobile UX comparison in Study II. }

After each condition, we asked participants to answer four 5-point Likert-scale questions on the presentation quality and comprehensibility.
We recorded participants' feedback on the Story presentation quality, summary, navigation, and consistency with the source Wikipedia article.

\subsubsection{Participants}
We invited participants via an internal mailing list of over 50 recipients in our organization. Eight people responded within the specified time frame, all were full-time employees in the U.S. We did not record demographic information due to organizational policy. Five indicated that they interacted with content in the Story format on a daily basis prior to the study, while the remaining three read Stories once a week. Each participant received a \${15} gift card or donation credit for their participation.

\begin{figure}[t]
  \centering
  \includegraphics[width=0.7\linewidth]{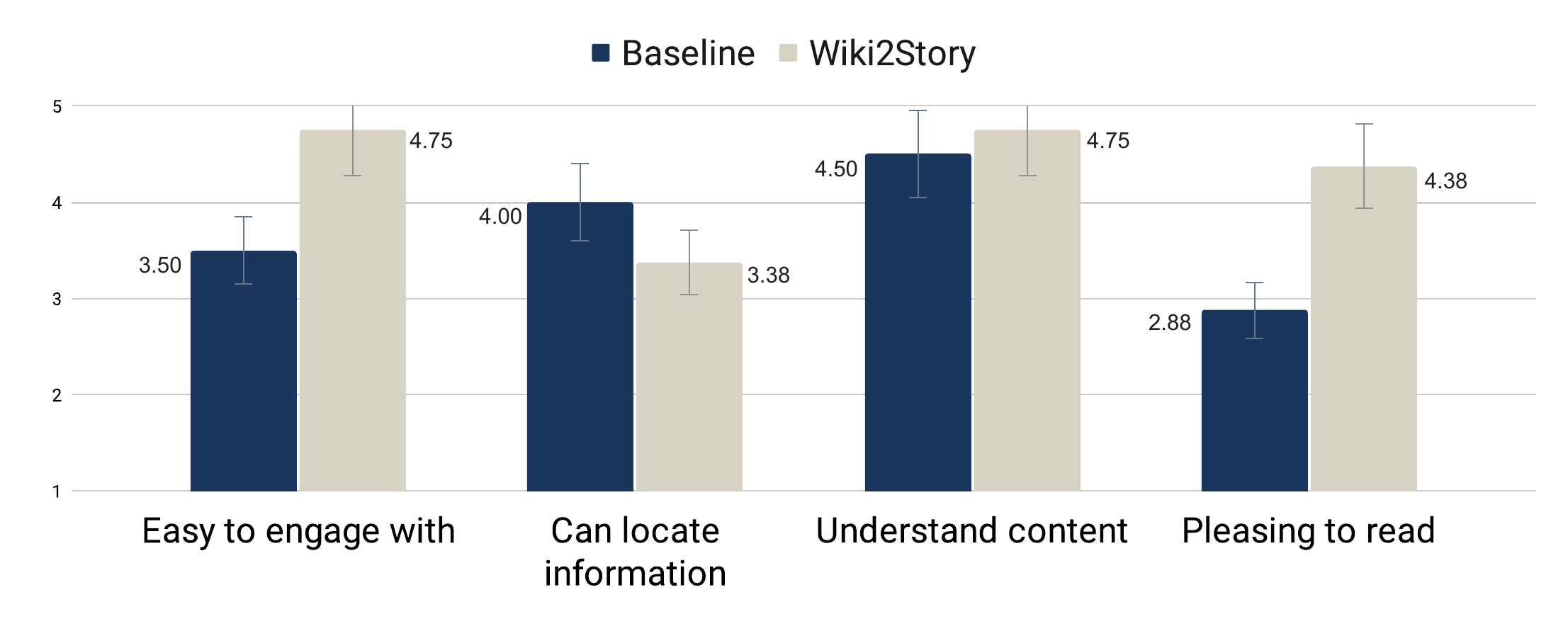}
  \caption{Participants' average response to five-point Likert-scale questions between Wiki2Story and the baseline in Study I. Each response was recorded after participants reviewed the task in the condition.}
  \Description{\systemname{} ...}
  \label{fig:ratings}
\end{figure}

\subsubsection{Results.}
Overall, the participants were satisfied with \systemname{}’s outputs (Median, $M = 4$, $\sigma = 0.5$), and preferred reading our Story than the source web article ($M = 4$, $\sigma = 0.8$). Participants highlighted \systemname{}’s strengths as, \userquote{I don't have to read all the details in the Wikipedia page, I think the facts of the article were well captured. Reading long Wikipedia articles is not fun. Being able to visually navigate the article is very great.} (P8), \userquote{This is a very compelling way to internalize information and to share that information with the people that you care about.} (P4), and \userquote{The Stories help me appreciate the images in the article better. I will not typically click on Wikipedia images to admire them.} (P1). Below we aggregate findings regarding the content summary and presentation. %(\systemname{} vs. the baseline, reading Wikipedia articles). \newline

\subheading{Content Extraction and Following.}
All participants suggested that our Stories facilitated their understanding of the topic, but no differently than the original article (Baseline: $M = 5$, $\sigma = 1.1$ vs. \systemname: $M = 5$, $\sigma = 0.7$). We found this encouraging as 
% our goal to present long text structural web articles into a Story would require 
we aim to ensure that our automatically-generated Stories could convey sufficient information for topic understanding.
Participants agreed that our Stories were pleasing to read ($M = 5$, $\sigma = 0.7$). In contrast, the baseline was described as \textit{``visually hard''} and \textit{``not fun''} with lower ratings ($M = 2.5$, $\sigma = 1.5$). 
P5 commented, \userquote{Personally, I don’t read Wikipedia articles because it is visually hard to read. I would rather glance at a snippet than open the entire article.}

Participants were positive on the quality of the summarization. They rated the summaries as helpful ($M = 5$, $\sigma = 0.7$) and accurate ($M = 4.5$, $\sigma = 1$). 
They described that the text-driven content of Wikipedia could be intimidating, while \systemname{}'s text snippets helped reduce cognitive strain. We observed that participants in the Baseline-first task ordering gave more positive feedback on the Story summaries.
% most likely due to their fresher memory of how dense articles can be.

% \subheading{Content Extraction.}
All eight participants agreed that the structure of the Stories were consistent with the baseline web article ($M = 5$, $\sigma = 0.7$). Participants in the Story-first task ordering particularly spoke of the navigation paths. 
Not going through the entire article was described as \textit{"very helpful"} and \textit{"clever"}. They mentioned that the navigation path allowed them to drill deeper into the sections of interests: %The navigation paths helped them feel like they are not missing anything; 
P4 noted, \userquote{I like how the navigation list lets me pick my own adventure of how to explore the article.} Participants agreed that the Story was easier to navigate than the baseline ($M = 4$, $\sigma = 1.1$).

\subheading{Visual-Based Presentation.}
All participants agreed that Wiki-2Story was engaging ($M = 5$, $\sigma = 0.5$) when compared to the Baseline ($M = 4$, $\sigma = 1.1$). Participants used words such as \userquote{fun}, \userquote{easy}, \userquote{visually appealing} to describe the Stories. The Story presentation helped participant pause and reflect on each page compared to simply scrolling up or down a web page. According to P7, \emph{``The Story feels more like a journey when you are going through it.''}
All the participants noted that \systemname{} increased their appreciation of Wikipedia images. Since every page had a leading image, participants were more likely to notice and comment on the image used within each page than when scrolling the Wikipedia page. Participants described the Story as \userquote{more visually appealing} (P4) and \userquote{bringing the images on the article to life} (P3). Participants described the equivalent process of savoring images on Wikipedia, either clicking on them to go to a new page or opening them in a new tab, as unpleasant. One participant highlighted the effectiveness of the image on the cover page for disambiguation - distinguishing between Wikipedia articles with similar titles.

% We suspect that The growth of the Story interface on several news and social platforms potentially played a role in facilitating user acceptance of the \systemname{} outputs. 

\subheading{Opportunities.}
P3 and P4, who indicated them as heavy mobile users, commented that Web Stories provided a more satisfying experience for mobile viewing. 
% Their familiarity with the Story format made it more likely that they would explore Wikipedia topics for fun, and for P4, even share them with friends and family.
% 
Several participants suggested that the Story format might be suitable for topic exploration instead of information seeking as a summary could lose details.
% use case is most beneficial for exploring a topic but not so much for looking for a specific detail. 
% This is due to the fact that, by definition, a Wikipedia section is typically a condense paragraph of facts and summarizing that paragraph, however helpful, 
% . While the Story presents an alternative mode for exploring articles, some use cases need all the details; 
P2 commented, \userquote{If you don't want to be overwhelmed, the Story is perhaps a good mode. However, the article might be preferred for people who want the details.}
% 
% Participants offered feedback on design improvement of our Story navigation. % Most participants mentioned that they were used to seeing the content tree on the web article right as they opened the page.
In addition, participants suggested making the end page available on every page for faster access.
% throughout the Story to allow easy in and out movements between section Stories. 
% 
% Furthermore, participants in the Wiki-first task ordering requested a search feature in a Story.
Although participants rated both conditions equally helped them locate information ($M = 4$, $\sigma = 0.9$ for \systemname{} and $M = 4$, $\sigma = 1.3$ for the baseline), several reflected their preferences for searching on the web article via a browser.
These findings align with prior work where a document is easier to scan through the content, while a videos or dynamic presentation is more engaging~\cite{Chi:2012:MAG:2380116.2380130,MakeupBreakdown_CHI21}. This suggests a need for enhancing the mobile browsing experience by more integration across the two.

%\subsection{Study II: Mobile Reading Experiences}
\mobilehci{\subsection{Study II: Mobile Reading Experiences}}
%To further investigate how viewers would value generated Stories on a mobile device outside a desktop context, we conducted a second study with participants who are self-rated as mobile heavy users (engage content in the Story format several times a day). 

\mobilehci{To further investigate how viewers would value generated Stories on a mobile device outside a desktop context, we conducted a second study with participants reviewing \systemname{} outputs on their mobile phones and comparing their reading experiences with the official Wikipedia iOS mobile application.}

\subsubsection{Study Design}
We selected four articles (including two from Study I) and defined two conditions: The \systemname{} condition was the automatically generated output of an article. The Baseline condition was the official Wikipedia iOS application\footnote{Wikipedia on the App Store, \url{https://apps.apple.com/us/app/wikipedia/id324715238}}, which presents an article's full content on a mobile-optimized view. \mobilehci{With the app version we tested in the study, we noticed that (1) quick facts were hidden by default, (2) images were positioned below a section, (3) each image was expanded to the screen width with inconsistent visual quality, and (4) all the presented content remained identical to the desktop version. Users could navigate by scrolling down to a section or tap the ``Contents'' side panel that lists the article's section and subsection titles (see Figure~\ref{fig:results} as examples).} 

We asked participants to review and describe the content with counterbalanced tasks and conditions.
% 
% \subsubsection{Participants}
We recruited six participants via a graduate student mailing list. \mobilehci{Participants were self-rated as heavy mobile device users (engage content in the Story format several times a day). They all used their personal iOS phone to complete Study II.} 

\subsubsection{Results.}
All the participants explored the selected topics in both formats and provided valuable feedback. Overall, our findings suggest that users favored the story format for exploring relatively new topics on Wikipedia when compared to the mobile app.  They commented on the benefits of the story format: \userquote{presenting articles as a Story will make it enjoyable to randomly explore new topics on my phone} (S1), \userquote{the image-first view adds a layer of interest to me. I've spent more time observing the images} (S3), and \userquote{this makes reading fun, it's like I'm on Instagram}(S4). Below, we summarize our findings: 

We observed strategies across participants for exploring topics using the different formats. In the mobile app baseline condition, participants typically scrolled through the article from top to bottom making random stops with their fingers at unspecified sections and taking a quick glance before scrolling again. When they used the Story format, they spent more time reading each of the text snippets in each Story page and observing the accompanying image on the page. 

Participants noted differences in their reading experience between both format. When asked to briefly discuss the article they had just explored, participants were more likely to recall what they had seen from the Story than the baseline condition. When probed on the reason they could recall the Story content better than baseline, participants stated that \userquote{the Story gives me a better overview of the topic. Though I will go to the original article in the mobile app if I wanted more details} (S2) and \userquote{I like that it slices the text into small digestible pieces} (S6).

\subsubsection{Other Remarks.}
Two of the participants mentioned that it will be great to carry over linked words from the article into the summary. This will allow the user to further explore Story formats of topics they do not know about linked from the current Story they are watching. Though we linked section summaries to their corresponding section in the article, this provides an opportunity to further allow users explore fresh topics. A participant remarked that the story format \userquote{brings the fact in my face}(S1) making it more likely to support further exploration.

\subsection{Study III: Informal Feedback from Designers}
To validate our results beyond general viewers, we invited three professional designers in our organization to share their Story creation processes and provide feedback on our outputs (the same Stories used in Study I) in a 30-minute session. Participants had similar experiences manually creating Stories using visual design tools. The designers highlighted that the visual appearance of a Story is the most critical goal in engaging readers: images should be high quality and properly cropped if necessary; a cover page should present a visually-appealing image; and content pages should maintain similar visual structures and tone. They advised that the visual layout should be diverse enough to retain a viewer's attention but not change too rapidly as to confuse them.

Designers appreciated an automatic tool to create a Story and thought the outputs were similar in quality to manually-created ones. They suggested that the system output could be a reasonable starting point for them to quickly modify details, such as adjusting or replacing images, experimenting different layouts, and modifying the font sizes or colors. They also suggested that where necessary, it might be justifiable to utilize quality images outside an article to match the content for a compelling Story.
% Moving the original implication section below and commented
\section{Discussion and Opportunities}

\subsection{Participant Feedback}
\mobilehci{In our two user studies testing \systemname{}'s outputs on desktop and mobile devices (compared with participant's desktop and mobile application versions of Wikipedia), we suggest that \systemname{}'s summary Stories were consistent with the source article and sufficient to facilitate user understanding of the article topic (Study I). \systemname{} leverages section-level extractive summarization and uses multiple Story levels to allow exploration of specific sections of the topic. 
Study II revealed an increased interest in a visual-first modality, which allows users to more critically examine images in Wikipedia articles. Participants noted that the Stories helped them stay engaged with the material when compared to the mobile application version. This suggests that presenting dense articles in digestible formats like a Web Story can help improve user retention of facts in the article compared to simply skimming the full article.}

\mobilehci{Overall, we found our study participants' feedback encouraging in terms of the quality and presentation of our automatically-generated Stories. Our approach extends beyond Wikipedia to convert hierarchical text articles to more contemporary formats such as Web Stories. Study participants were positive about \systemname{}'s ability to make dense content more accessible. A visual-first modality that allows for dynamic exploration of the source article can support people who do not understand the English language (Wikipedia's dominant language), or people who have special visual or reading needs which go often undiagnosed in many parts of the world. The Story format offers a compelling mode for addressing this need especially for mobile-first users. Our study results show that \systemname{} users valued the power of Web Stories to bring text articles to life in an engaging and visually-appealing way. Since users will still find use for the full article, \systemname{} does not attempt to replace full articles but provide an additional format for engaging with them.}

\subsection{Design Implications}
\mobilehci{Our analysis of 55 Web Stories and their corresponding source articles offers some insights into the manual labor involved in the design and creation of Web Stories. To create a Web Story from a source article, the designer needs to establish salient points from the article to be summarized in the story, identify what sections and paragraphs to include, characterize the interplay with source article length and structure and Story length and structure, and apply visuals from the article that help communicate the message in an accurate and engaging manner. Our research asks the question: how well can we automatically replicate the laborious process of Web Story creation from dense text articles in an accurate, efficient, and scalable manner? In answering this question, we note that while the average Wikipedia article tends to be comprehensive and text-heavy, well-crafted Web Stories are meant to be concise and visually-appealing.The challenge is to effectively constrain the summarization to retain consistency and correctness of the output with respect to the source article while keeping a user engaged using the Web Story interface.}

\mobilehci{\systemname{} is designed to automatically generate visually-appealing digestible Web Stories from dense hierarchical Wikipedia articles. We contribute an automatic approach to multimodal summarization at scale that is applicable by users for quickly learning about new topics, and carried out studies to test whether they found this approach useful and exciting. \systemname{} builds on prior works in multimodal summarization~\cite{zhu2018msmo, jangra2021survey} that have explored approaches for summarizing text materials and images and output summarized text with aligned images. We extend these works to structured articles where maintaining article hierarchy and facts matter; and presenting outputs in a novel Web Story format that supports accessibility, interoperability, and scalability.}

\mobilehci{Platforms that generate dense text articles such as Wikipedia typically have mobile application and responsive web application versions of their content. These mobile adapted versions help present content in flexible layouts fit for mobile viewing~\cite{gardner2011responsive, mohorovivcic2013implementing}. However, our results show that mobile users did not find the reading experience pleasurable nor did they find the content easy to remember when compared to the Story format. Prior work have shown that it is easier for readers to glean information from images + text than text alone~\cite{li2017multi}. For their multimodal summarization system, MSMO, Zhu et al.~\cite{zhu2018msmo} report that images and text work effectively together as a single output, with images helping users grasp events, and texts providing more context related to the event. Our findings provide evidence that the Web Story format is effectively suited for bringing visuals to the forefront of a user's experience while providing accompanying text for reference.}

\mobilehci{Our results also show that each end-to-end Web Story generation process took between 2 and 5 minutes to complete. Though related works such as Doc2PTT~\cite{fu2022doc2ppt} and MSMO~\cite{zhu2018msmo} do not mention end-to-end generation times, scaling \systemname{} to millions of article requires extra consideration to be made into how parts of the pipeline play a role in Story generation. A major part of \systemname{}'s short generation turnaround is attributable to the lightweight nature of the Web Story HTML framework used for its output modality. This makes it easy to scale to hundreds of millions of articles and to create new or refresh updated articles on the fly. Web Stories offer the creative and visual benefits of formats such as video and slideshows but without the added production and streaming overload since it requires less software and hardware resources to maintain. }

\subsection{Limitations}
Below, we highlight limitations of \systemname{}'s current pipeline and discuss potential opportunities and insights for future work.

\subheading{Expand supplementary visuals.} 
We learned from our studies that the Web Story format guided users to focus on the leading images. \systemname{} reuses images in cases where the number of content sections is orders of magnitude more than the available images on the page. Though we presently restrict our input pages to those with at least five images, this reuse of images may cause repetitions and impact a Story's visual appeal. \mobilehci{Future work may consider opportunities of searching for supplementary images \cite{VisualCaptions_CHI23}), generating images ~\cite{ImagenEditor_CVPR23} or videos~\cite{ImagenVideo_CVPR22}, or creating placeholders, to complement the accompanying summary text in a Story page. We also look forward to the future where content transformation will gradually encourage community contributors to include more visuals as part of their articles to support users who prefer visual learning. }
% 
%Beyond selecting templates based on text length, future work could perform saliency detection on the input image or learn from the summary text to determine the best template for a given card. 
There is also an avenue for exploring image presentation options, such as creating collages for the cover page using constituent images in the article or present in a landscape ratio aspect, that we leave it to future work.

% \subheading{Multi-level summarization.} 
% A participant might describe a story as concise with sufficient detail while another participant wishes they could read more within the Story. One naive solution is to link the web article from the story and allow users go to the web article should they wish to see the full details. A more robust
% approach might involve improving \systemname{}'s summarization module to include short, medium, and longer modes. Users can then select whichever mode they prefer and toggle this setting as they wish. \newline

\subheading{Support personalization and multilinguality.} 
We observed in the study that users showed varying preferences on the amount of content details in a Story. 
Future work can support personalization for different needs and use cases. Finally, 
% Tens of millions of Wikipedia articles have been written in more than 260 languages\footnote{https://en.wikipedia.org/wiki/Wikipedia:Multilingual\_statistics}. For context, only 11\% of the over 58 million published Wikipedia articles are in English\footnote{https://en.wikipedia.org/wiki/List\_of\_Wikipedias}. To reach a wider audience, Wiki2Story's pipeline can be 
we aim to extend our pipeline to support multiple languages. 
A large amount of Wikipedia articles are authored in different languages\footnote{As of September 2022, there are over 329 languages in the Wikipedia community. Source: \url{https://meta.wikimedia.org/wiki/List_of_Wikipedias}}
While Wikipedia articles do not necessarily share the same content across languages~\cite{bruckman2022should}, a possible approach is to leverage progress in language translation techniques~\cite{Translation_Jia2019} to translate Stories from one language to another. This is an exciting future direction to support our goal of making content accessible across reading preferences (from in-depth to an overview), device forms (from desktop to mobile), and languages.
% Significant progress in this direction will be driven by progress in NLP research for multilingual text summarization and cross-cultural user experience studies. 

\subheading{Enable interactive editing.} While we provide a review tool to inspect the content mapping between a generated Story and the source article (see Figure~\ref{fig:review_ui}), we suggest future supports to jumpstart a Story authoring process for content creator. Authors could guide AI to identify alternatives, re-layouts, and rephrasing within the context\mobilehci{, similar to recent generative image efforts~\cite{ImagenEditor_CVPR23}}. By learning from human editing decisions, a tool could better capture the styles and languages for Story consumption, which we leave for future opportunities.

\section{Conclusion}
A vast amount of online materials are created as lengthy structured articles which may not be suitable for users who prefer more visual, mobile-first, and faster-access modalities. We presented \systemname{}, an automatic approach that converts a structural article to a multi-path interactive Web Story that is visually aesthetic and easily digestible by a viewer. Our evaluation showed that computational approaches to extract, summarize, and visualize content as interactive Stories could support user engagement and comprehension. % with an article and their comprehension of the content. 
This approach could potentially be applied to support content accessibility for a wide range of structured documents beyond Wikipedia.

\begin{acks}
We thank all the participants in our studies for their valuable insight in moving this research forward.
\end{acks}

\bibliographystyle{ACM-Reference-Format}
\bibliography{0_main}

%%% -*-BibTeX-*-
%%% Do NOT edit. File created by BibTeX with style
%%% ACM-Reference-Format-Journals [18-Jan-2012].

\begin{thebibliography}{81}

%%% ====================================================================
%%% NOTE TO THE USER: you can override these defaults by providing
%%% customized versions of any of these macros before the \bibliography
%%% command.  Each of them MUST provide its own final punctuation,
%%% except for \shownote{}, \showDOI{}, and \showURL{}.  The latter two
%%% do not use final punctuation, in order to avoid confusing it with
%%% the Web address.
%%%
%%% To suppress output of a particular field, define its macro to expand
%%% to an empty string, or better, \unskip, like this:
%%%
%%% \newcommand{\showDOI}[1]{\unskip}   % LaTeX syntax
%%%
%%% \def \showDOI #1{\unskip}           % plain TeX syntax
%%%
%%% ====================================================================

\ifx \showCODEN    \undefined \def \showCODEN     #1{\unskip}     \fi
\ifx \showDOI      \undefined \def \showDOI       #1{#1}\fi
\ifx \showISBNx    \undefined \def \showISBNx     #1{\unskip}     \fi
\ifx \showISBNxiii \undefined \def \showISBNxiii  #1{\unskip}     \fi
\ifx \showISSN     \undefined \def \showISSN      #1{\unskip}     \fi
\ifx \showLCCN     \undefined \def \showLCCN      #1{\unskip}     \fi
\ifx \shownote     \undefined \def \shownote      #1{#1}          \fi
\ifx \showarticletitle \undefined \def \showarticletitle #1{#1}   \fi
\ifx \showURL      \undefined \def \showURL       {\relax}        \fi
% The following commands are used for tagged output and should be
% invisible to TeX
\providecommand\bibfield[2]{#2}
\providecommand\bibinfo[2]{#2}
\providecommand\natexlab[1]{#1}
\providecommand\showeprint[2][]{arXiv:#2}

\bibitem[\protect\citeauthoryear{Allahyari, Pouriyeh, Assefi, Safaei, Trippe,
  Gutierrez, and Kochut}{Allahyari et~al\mbox{.}}{2017}]%
        {TextSummarizationSurvey2017}
\bibfield{author}{\bibinfo{person}{Mehdi Allahyari}, \bibinfo{person}{Seyedamin
  Pouriyeh}, \bibinfo{person}{Mehdi Assefi}, \bibinfo{person}{Saeid Safaei},
  \bibinfo{person}{Elizabeth~D Trippe}, \bibinfo{person}{Juan~B Gutierrez},
  {and} \bibinfo{person}{Krys Kochut}.} \bibinfo{year}{2017}\natexlab{}.
\newblock \showarticletitle{Text summarization techniques: a brief survey}.
\newblock \bibinfo{journal}{\emph{arXiv preprint arXiv:1707.02268}}
  (\bibinfo{year}{2017}).
\newblock


\bibitem[\protect\citeauthoryear{Amitay and Paris}{Amitay and Paris}{2000}]%
        {WebSiteSummarization_ICIKM2000}
\bibfield{author}{\bibinfo{person}{Einat Amitay} {and}
  \bibinfo{person}{C{\'e}cile Paris}.} \bibinfo{year}{2000}\natexlab{}.
\newblock \showarticletitle{Automatically summarising web sites: is there a way
  around it?}. In \bibinfo{booktitle}{\emph{Proceedings of the ninth
  international conference on Information and knowledge management}}.
  \bibinfo{pages}{173--179}.
\newblock


\bibitem[\protect\citeauthoryear{AMP}{AMP}{2022a}]%
        {WebStoriesPracticesAMP}
\bibfield{author}{\bibinfo{person}{AMP}.} \bibinfo{year}{2022}\natexlab{a}.
\newblock \bibinfo{booktitle}{\emph{Best practices for creating a successful
  Web Story}}.
\newblock
\urldef\tempurl%
\url{https://amp.dev/documentation/guides-and-tutorials/start/create_successful_stories/?format=websites}
\showURL{%
Retrieved April, 2022 from \tempurl}


\bibitem[\protect\citeauthoryear{AMP}{AMP}{2022b}]%
        {WebStoriesAMP}
\bibfield{author}{\bibinfo{person}{AMP}.} \bibinfo{year}{2022}\natexlab{b}.
\newblock \bibinfo{booktitle}{\emph{Web Stories - amp.dev}}.
\newblock
\urldef\tempurl%
\url{https://amp.dev/about/stories/}
\showURL{%
Retrieved April, 2022 from \tempurl}


\bibitem[\protect\citeauthoryear{Arroyo, Postels, and Tombari}{Arroyo
  et~al\mbox{.}}{2021}]%
        {arroyo2021variational}
\bibfield{author}{\bibinfo{person}{Diego~Martin Arroyo}, \bibinfo{person}{Janis
  Postels}, {and} \bibinfo{person}{Federico Tombari}.}
  \bibinfo{year}{2021}\natexlab{}.
\newblock \showarticletitle{Variational transformer networks for layout
  generation}. In \bibinfo{booktitle}{\emph{Proceedings of the IEEE/CVF
  Conference on Computer Vision and Pattern Recognition}}.
  \bibinfo{pages}{13642--13652}.
\newblock


\bibitem[\protect\citeauthoryear{Baxendale}{Baxendale}{1958}]%
        {baxendale1958machine}
\bibfield{author}{\bibinfo{person}{Phyllis~B Baxendale}.}
  \bibinfo{year}{1958}\natexlab{}.
\newblock \showarticletitle{Machine-made index for technical literature—an
  experiment}.
\newblock \bibinfo{journal}{\emph{IBM Journal of research and development}}
  \bibinfo{volume}{2}, \bibinfo{number}{4} (\bibinfo{year}{1958}),
  \bibinfo{pages}{354--361}.
\newblock


\bibitem[\protect\citeauthoryear{Bergman, Castelli, Lau, and Oblinger}{Bergman
  et~al\mbox{.}}{2005}]%
        {DocWizards_UIST05}
\bibfield{author}{\bibinfo{person}{Lawrence Bergman}, \bibinfo{person}{Vittorio
  Castelli}, \bibinfo{person}{Tessa Lau}, {and} \bibinfo{person}{Daniel
  Oblinger}.} \bibinfo{year}{2005}\natexlab{}.
\newblock \showarticletitle{DocWizards: A System for Authoring Follow-Me
  Documentation Wizards}. In \bibinfo{booktitle}{\emph{Proceedings of the 18th
  Annual ACM Symposium on User Interface Software and Technology}} (Seattle,
  WA, USA) \emph{(\bibinfo{series}{UIST '05})}. \bibinfo{publisher}{Association
  for Computing Machinery}, \bibinfo{address}{New York, NY, USA},
  \bibinfo{pages}{191–200}.
\newblock
\showISBNx{1595932712}
\urldef\tempurl%
\url{https://doi.org/10.1145/1095034.1095067}
\showDOI{\tempurl}


\bibitem[\protect\citeauthoryear{Bian, Yang, and Chua}{Bian
  et~al\mbox{.}}{2013}]%
        {bian2013multimedia}
\bibfield{author}{\bibinfo{person}{Jingwen Bian}, \bibinfo{person}{Yang Yang},
  {and} \bibinfo{person}{Tat-Seng Chua}.} \bibinfo{year}{2013}\natexlab{}.
\newblock \showarticletitle{Multimedia summarization for trending topics in
  microblogs}. In \bibinfo{booktitle}{\emph{Proceedings of the 22nd ACM
  international Conference on information \& knowledge management}}.
  \bibinfo{pages}{1807--1812}.
\newblock


\bibitem[\protect\citeauthoryear{Bruckman}{Bruckman}{2022}]%
        {bruckman2022should}
\bibfield{author}{\bibinfo{person}{Amy~S Bruckman}.}
  \bibinfo{year}{2022}\natexlab{}.
\newblock \bibinfo{booktitle}{\emph{Should You Believe Wikipedia?: Online
  Communities and the Construction of Knowledge}}.
\newblock \bibinfo{publisher}{Cambridge University Press}.
\newblock


\bibitem[\protect\citeauthoryear{Chen, Xiao, and Gao}{Chen
  et~al\mbox{.}}{2010}]%
        {iSlideShow2010}
\bibfield{author}{\bibinfo{person}{Jiajian Chen}, \bibinfo{person}{Jun Xiao},
  {and} \bibinfo{person}{Yuli Gao}.} \bibinfo{year}{2010}\natexlab{}.
\newblock \showarticletitle{ISlideShow: A Content-Aware Slideshow System}. In
  \bibinfo{booktitle}{\emph{Proceedings of the 15th International Conference on
  Intelligent User Interfaces}} (Hong Kong, China) \emph{(\bibinfo{series}{IUI
  '10})}. \bibinfo{publisher}{Association for Computing Machinery},
  \bibinfo{address}{New York, NY, USA}, \bibinfo{pages}{293--296}.
\newblock
\showISBNx{9781605585154}
\urldef\tempurl%
\url{https://doi.org/10.1145/1719970.1720014}
\showDOI{\tempurl}


\bibitem[\protect\citeauthoryear{Chen, Chu, Kuo, Weng, and Wu}{Chen
  et~al\mbox{.}}{2006}]%
        {TilingSlideshow2006}
\bibfield{author}{\bibinfo{person}{Jun-Cheng Chen}, \bibinfo{person}{Wei-Ta
  Chu}, \bibinfo{person}{Jin-Hau Kuo}, \bibinfo{person}{Chung-Yi Weng}, {and}
  \bibinfo{person}{Ja-Ling Wu}.} \bibinfo{year}{2006}\natexlab{}.
\newblock \showarticletitle{Tiling Slideshow}. In
  \bibinfo{booktitle}{\emph{Proceedings of the 14th ACM International
  Conference on Multimedia}} (Santa Barbara, CA, USA)
  \emph{(\bibinfo{series}{MM '06})}. \bibinfo{publisher}{Association for
  Computing Machinery}, \bibinfo{address}{New York, NY, USA},
  \bibinfo{pages}{25--34}.
\newblock
\showISBNx{1595934472}
\urldef\tempurl%
\url{https://doi.org/10.1145/1180639.1180653}
\showDOI{\tempurl}


\bibitem[\protect\citeauthoryear{Chi, Dong, Frueh, Colonna, Kwatra, and
  Essa}{Chi et~al\mbox{.}}{2022}]%
        {doc2video_uist22}
\bibfield{author}{\bibinfo{person}{Peggy Chi}, \bibinfo{person}{Tao Dong},
  \bibinfo{person}{Christian Frueh}, \bibinfo{person}{Brian Colonna},
  \bibinfo{person}{Vivek Kwatra}, {and} \bibinfo{person}{Irfan Essa}.}
  \bibinfo{year}{2022}\natexlab{}.
\newblock \showarticletitle{Synthesis-Assisted Video Prototyping From a
  Document}. In \bibinfo{booktitle}{\emph{Proceedings of the 35th Annual ACM
  Symposium on User Interface Software and Technology}} (Bend, OR, USA)
  \emph{(\bibinfo{series}{UIST '22})}. \bibinfo{publisher}{Association for
  Computing Machinery}, \bibinfo{address}{New York, NY, USA}, Article
  \bibinfo{articleno}{16}, \bibinfo{numpages}{10}~pages.
\newblock
\showISBNx{9781450393201}
\urldef\tempurl%
\url{https://doi.org/10.1145/3526113.3545676}
\showDOI{\tempurl}


\bibitem[\protect\citeauthoryear{Chi, Frey, Panovich, and Essa}{Chi
  et~al\mbox{.}}{2021}]%
        {HowToCut_UIST21}
\bibfield{author}{\bibinfo{person}{Peggy Chi}, \bibinfo{person}{Nathan Frey},
  \bibinfo{person}{Katrina Panovich}, {and} \bibinfo{person}{Irfan Essa}.}
  \bibinfo{year}{2021}\natexlab{}.
\newblock \showarticletitle{Automatic Instructional Video Creation from a
  Markdown-Formatted Tutorial}. In \bibinfo{booktitle}{\emph{The 34th Annual
  ACM Symposium on User Interface Software and Technology}} (Virtual Event,
  USA) \emph{(\bibinfo{series}{UIST '21})}. \bibinfo{publisher}{Association for
  Computing Machinery}, \bibinfo{address}{New York, NY, USA},
  \bibinfo{pages}{677–690}.
\newblock
\showISBNx{9781450386357}
\urldef\tempurl%
\url{https://doi.org/10.1145/3472749.3474778}
\showDOI{\tempurl}


\bibitem[\protect\citeauthoryear{Chi, Sun, Panovich, and Essa}{Chi
  et~al\mbox{.}}{2020}]%
        {URL2Video_UIST20}
\bibfield{author}{\bibinfo{person}{Peggy Chi}, \bibinfo{person}{Zheng Sun},
  \bibinfo{person}{Katrina Panovich}, {and} \bibinfo{person}{Irfan Essa}.}
  \bibinfo{year}{2020}\natexlab{}.
\newblock \showarticletitle{Automatic Video Creation From a Web Page}. In
  \bibinfo{booktitle}{\emph{Proceedings of the 33rd Annual ACM Symposium on
  User Interface Software and Technology}} (Virtual Event, USA)
  \emph{(\bibinfo{series}{UIST '20})}. \bibinfo{publisher}{Association for
  Computing Machinery}, \bibinfo{address}{New York, NY, USA},
  \bibinfo{pages}{279–292}.
\newblock
\showISBNx{9781450375146}
\urldef\tempurl%
\url{https://doi.org/10.1145/3379337.3415814}
\showDOI{\tempurl}


\bibitem[\protect\citeauthoryear{Chi, Ahn, Ren, Dontcheva, Li, and
  Hartmann}{Chi et~al\mbox{.}}{2012}]%
        {Chi:2012:MAG:2380116.2380130}
\bibfield{author}{\bibinfo{person}{Pei-Yu Chi}, \bibinfo{person}{Sally Ahn},
  \bibinfo{person}{Amanda Ren}, \bibinfo{person}{Mira Dontcheva},
  \bibinfo{person}{Wilmot Li}, {and} \bibinfo{person}{Bj\"{o}rn Hartmann}.}
  \bibinfo{year}{2012}\natexlab{}.
\newblock \showarticletitle{MixT: Automatic Generation of Step-by-step Mixed
  Media Tutorials}. In \bibinfo{booktitle}{\emph{Proceedings of the 25th Annual
  ACM Symposium on User Interface Software and Technology}} (Cambridge,
  Massachusetts, USA) \emph{(\bibinfo{series}{UIST '12})}.
  \bibinfo{publisher}{ACM}, \bibinfo{address}{New York, NY, USA},
  \bibinfo{pages}{93--102}.
\newblock
\showISBNx{978-1-4503-1580-7}
\urldef\tempurl%
\url{https://doi.org/10.1145/2380116.2380130}
\showDOI{\tempurl}


\bibitem[\protect\citeauthoryear{Chu and Kao}{Chu and Kao}{2017}]%
        {chu2017blog}
\bibfield{author}{\bibinfo{person}{Wei-Ta Chu} {and} \bibinfo{person}{Ming-Chih
  Kao}.} \bibinfo{year}{2017}\natexlab{}.
\newblock \showarticletitle{Blog article summarization with image-text
  alignment techniques}. In \bibinfo{booktitle}{\emph{2017 IEEE International
  Symposium on Multimedia (ISM)}}. IEEE, \bibinfo{pages}{244--247}.
\newblock


\bibitem[\protect\citeauthoryear{Chuang and Yang}{Chuang and Yang}{2000}]%
        {chuang2000extracting}
\bibfield{author}{\bibinfo{person}{Wesley~T Chuang} {and}
  \bibinfo{person}{Jihoon Yang}.} \bibinfo{year}{2000}\natexlab{}.
\newblock \showarticletitle{Extracting sentence segments for text
  summarization: a machine learning approach}. In
  \bibinfo{booktitle}{\emph{Proceedings of the 23rd annual international ACM
  SIGIR conference on Research and development in information retrieval}}.
  \bibinfo{pages}{152--159}.
\newblock


\bibitem[\protect\citeauthoryear{Delort, Bouchon-Meunier, and Rifqi}{Delort
  et~al\mbox{.}}{2003}]%
        {WebDocSummarization_ACMHypertext2003}
\bibfield{author}{\bibinfo{person}{J-Y Delort}, \bibinfo{person}{Bernadette
  Bouchon-Meunier}, {and} \bibinfo{person}{Maria Rifqi}.}
  \bibinfo{year}{2003}\natexlab{}.
\newblock \showarticletitle{Enhanced web document summarization using
  hyperlinks}. In \bibinfo{booktitle}{\emph{Proceedings of the fourteenth ACM
  conference on Hypertext and hypermedia}}. \bibinfo{pages}{208--215}.
\newblock


\bibitem[\protect\citeauthoryear{Dong, Yang, Wang, Wei, Liu, Wang, Gao, Zhou,
  and Hon}{Dong et~al\mbox{.}}{2019}]%
        {dong2019unified}
\bibfield{author}{\bibinfo{person}{Li Dong}, \bibinfo{person}{Nan Yang},
  \bibinfo{person}{Wenhui Wang}, \bibinfo{person}{Furu Wei},
  \bibinfo{person}{Xiaodong Liu}, \bibinfo{person}{Yu Wang},
  \bibinfo{person}{Jianfeng Gao}, \bibinfo{person}{Ming Zhou}, {and}
  \bibinfo{person}{Hsiao-Wuen Hon}.} \bibinfo{year}{2019}\natexlab{}.
\newblock \showarticletitle{Unified language model pre-training for natural
  language understanding and generation}.
\newblock \bibinfo{journal}{\emph{Advances in Neural Information Processing
  Systems}}  \bibinfo{volume}{32} (\bibinfo{year}{2019}).
\newblock


\bibitem[\protect\citeauthoryear{Edmundson}{Edmundson}{1969}]%
        {edmundson1969new}
\bibfield{author}{\bibinfo{person}{Harold~P Edmundson}.}
  \bibinfo{year}{1969}\natexlab{}.
\newblock \showarticletitle{New methods in automatic extracting}.
\newblock \bibinfo{journal}{\emph{Journal of the ACM (JACM)}}
  \bibinfo{volume}{16}, \bibinfo{number}{2} (\bibinfo{year}{1969}),
  \bibinfo{pages}{264--285}.
\newblock


\bibitem[\protect\citeauthoryear{Fang and Zhang}{Fang and Zhang}{2017}]%
        {fang2017creatism}
\bibfield{author}{\bibinfo{person}{Hui Fang} {and} \bibinfo{person}{Meng
  Zhang}.} \bibinfo{year}{2017}\natexlab{}.
\newblock \showarticletitle{Creatism: A deep-learning photographer capable of
  creating professional work}.
\newblock \bibinfo{journal}{\emph{arXiv preprint arXiv:1707.03491}}
  (\bibinfo{year}{2017}).
\newblock


\bibitem[\protect\citeauthoryear{Fiorella and Mayer}{Fiorella and
  Mayer}{2018}]%
        {fiorella2018works}
\bibfield{author}{\bibinfo{person}{Logan Fiorella} {and}
  \bibinfo{person}{Richard~E Mayer}.} \bibinfo{year}{2018}\natexlab{}.
\newblock \bibinfo{title}{What works and doesn't work with instructional
  video}.
\newblock , \bibinfo{numpages}{465--470}~pages.
\newblock


\bibitem[\protect\citeauthoryear{Foundation}{Foundation}{2022}]%
        {VideoWiki}
\bibfield{author}{\bibinfo{person}{Wikimedia Foundation}.}
  \bibinfo{year}{2022}\natexlab{}.
\newblock \bibinfo{booktitle}{\emph{VideoWiki}}.
\newblock
\urldef\tempurl%
\url{https://meta.wikimedia.org/wiki/VideoWiki}
\showURL{%
Retrieved November, 2022 from \tempurl}


\bibitem[\protect\citeauthoryear{Fraser, Ngoon, Dontcheva, and Klemmer}{Fraser
  et~al\mbox{.}}{2019}]%
        {Replay_CHI2019}
\bibfield{author}{\bibinfo{person}{C.~Ailie Fraser}, \bibinfo{person}{Tricia~J.
  Ngoon}, \bibinfo{person}{Mira Dontcheva}, {and} \bibinfo{person}{Scott
  Klemmer}.} \bibinfo{year}{2019}\natexlab{}.
\newblock \showarticletitle{RePlay: Contextually Presenting Learning Videos
  Across Software Applications}. In \bibinfo{booktitle}{\emph{Proceedings of
  the 2019 CHI Conference on Human Factors in Computing Systems}} (Glasgow,
  Scotland Uk) \emph{(\bibinfo{series}{CHI '19})}.
  \bibinfo{publisher}{Association for Computing Machinery},
  \bibinfo{address}{New York, NY, USA}, \bibinfo{pages}{1–13}.
\newblock
\showISBNx{9781450359702}
\urldef\tempurl%
\url{https://doi.org/10.1145/3290605.3300527}
\showDOI{\tempurl}


\bibitem[\protect\citeauthoryear{Fu, Wang, McDuff, and Song}{Fu
  et~al\mbox{.}}{2022}]%
        {fu2022doc2ppt}
\bibfield{author}{\bibinfo{person}{Tsu-Jui Fu}, \bibinfo{person}{William~Yang
  Wang}, \bibinfo{person}{Daniel McDuff}, {and} \bibinfo{person}{Yale Song}.}
  \bibinfo{year}{2022}\natexlab{}.
\newblock \showarticletitle{Doc2ppt: Automatic presentation slides generation
  from scientific documents}. In \bibinfo{booktitle}{\emph{Proceedings of the
  AAAI Conference on Artificial Intelligence}}, Vol.~\bibinfo{volume}{36}.
  \bibinfo{pages}{634--642}.
\newblock


\bibitem[\protect\citeauthoryear{Gardner}{Gardner}{2011}]%
        {gardner2011responsive}
\bibfield{author}{\bibinfo{person}{Brett~S Gardner}.}
  \bibinfo{year}{2011}\natexlab{}.
\newblock \showarticletitle{Responsive web design: Enriching the user
  experience}.
\newblock \bibinfo{journal}{\emph{Sigma Journal: Inside the Digital Ecosystem}}
  \bibinfo{volume}{11}, \bibinfo{number}{1} (\bibinfo{year}{2011}),
  \bibinfo{pages}{13--19}.
\newblock


\bibitem[\protect\citeauthoryear{Google}{Google}{2022a}]%
        {GoogleConsumerInsights}
\bibfield{author}{\bibinfo{person}{Google}.} \bibinfo{year}{2022}\natexlab{a}.
\newblock \bibinfo{booktitle}{\emph{Think with Google: Consumer Insights}}.
\newblock
\urldef\tempurl%
\url{https://www.thinkwithgoogle.com/consumer-insights/consumer-trends/millennial-learning-statistics/}
\showURL{%
Retrieved April, 2022 from \tempurl}


\bibitem[\protect\citeauthoryear{Google}{Google}{2022b}]%
        {WebStoriesEditor}
\bibfield{author}{\bibinfo{person}{Google}.} \bibinfo{year}{2022}\natexlab{b}.
\newblock \bibinfo{booktitle}{\emph{Tools - Web Stories on Google}}.
\newblock
\urldef\tempurl%
\url{https://stories.google/tools/}
\showURL{%
Retrieved April, 2022 from \tempurl}


\bibitem[\protect\citeauthoryear{Hermann, Kocisky, Grefenstette, Espeholt, Kay,
  Suleyman, and Blunsom}{Hermann et~al\mbox{.}}{2015}]%
        {NIPS2015_afdec700}
\bibfield{author}{\bibinfo{person}{Karl~Moritz Hermann}, \bibinfo{person}{Tomas
  Kocisky}, \bibinfo{person}{Edward Grefenstette}, \bibinfo{person}{Lasse
  Espeholt}, \bibinfo{person}{Will Kay}, \bibinfo{person}{Mustafa Suleyman},
  {and} \bibinfo{person}{Phil Blunsom}.} \bibinfo{year}{2015}\natexlab{}.
\newblock \showarticletitle{Teaching Machines to Read and Comprehend}. In
  \bibinfo{booktitle}{\emph{Advances in Neural Information Processing
  Systems}}, \bibfield{editor}{\bibinfo{person}{C.~Cortes},
  \bibinfo{person}{N.~Lawrence}, \bibinfo{person}{D.~Lee},
  \bibinfo{person}{M.~Sugiyama}, {and} \bibinfo{person}{R.~Garnett}} (Eds.),
  Vol.~\bibinfo{volume}{28}. \bibinfo{publisher}{Curran Associates, Inc.}
\newblock
\urldef\tempurl%
\url{https://proceedings.neurips.cc/paper/2015/file/afdec7005cc9f14302cd0474fd0f3c96-Paper.pdf}
\showURL{%
\tempurl}


\bibitem[\protect\citeauthoryear{Ho, Chan, Saharia, Whang, Gao, Gritsenko,
  Kingma, Poole, Norouzi, Fleet, and Salimans}{Ho et~al\mbox{.}}{2022}]%
        {ImagenVideo_CVPR22}
\bibfield{author}{\bibinfo{person}{Jonathan Ho}, \bibinfo{person}{William
  Chan}, \bibinfo{person}{Chitwan Saharia}, \bibinfo{person}{Jay Whang},
  \bibinfo{person}{Ruiqi Gao}, \bibinfo{person}{Alexey Gritsenko},
  \bibinfo{person}{Diederik~P. Kingma}, \bibinfo{person}{Ben Poole},
  \bibinfo{person}{Mohammad Norouzi}, \bibinfo{person}{David~J. Fleet}, {and}
  \bibinfo{person}{Tim Salimans}.} \bibinfo{year}{2022}\natexlab{}.
\newblock \showarticletitle{Imagen Video: High Definition Video Generation with
  Diffusion Models}. In \bibinfo{booktitle}{\emph{CVPR}}.
\newblock
\showeprint{2210.02303}


\bibitem[\protect\citeauthoryear{Hovy, Lin, et~al\mbox{.}}{Hovy
  et~al\mbox{.}}{1999}]%
        {hovy1999automated}
\bibfield{author}{\bibinfo{person}{Eduard Hovy}, \bibinfo{person}{Chin-Yew
  Lin}, {et~al\mbox{.}}} \bibinfo{year}{1999}\natexlab{}.
\newblock \showarticletitle{Automated text summarization in SUMMARIST}.
\newblock \bibinfo{journal}{\emph{Advances in automatic text summarization}}
  \bibinfo{volume}{14} (\bibinfo{year}{1999}), \bibinfo{pages}{81--94}.
\newblock


\bibitem[\protect\citeauthoryear{Inc}{Inc}{2022}]%
        {SpeechAPI}
\bibfield{author}{\bibinfo{person}{Google Inc}.}
  \bibinfo{year}{2022}\natexlab{}.
\newblock \bibinfo{booktitle}{\emph{Text-to-Speech: Lifelike Speech
  Synthesis}}.
\newblock
\urldef\tempurl%
\url{https://cloud.google.com/text-to-speech/}
\showURL{%
Retrieved April, 2022 from \tempurl}


\bibitem[\protect\citeauthoryear{Jangra, Mukherjee, Jatowt, Saha, and
  Hasanuzzaman}{Jangra et~al\mbox{.}}{2021}]%
        {jangra2021survey}
\bibfield{author}{\bibinfo{person}{Anubhav Jangra}, \bibinfo{person}{Sourajit
  Mukherjee}, \bibinfo{person}{Adam Jatowt}, \bibinfo{person}{Sriparna Saha},
  {and} \bibinfo{person}{Mohammad Hasanuzzaman}.}
  \bibinfo{year}{2021}\natexlab{}.
\newblock \showarticletitle{A survey on multi-modal summarization}.
\newblock \bibinfo{journal}{\emph{Comput. Surveys}} (\bibinfo{year}{2021}).
\newblock


\bibitem[\protect\citeauthoryear{Jasti}{Jasti}{2022}]%
        {WebStoriesStats}
\bibfield{author}{\bibinfo{person}{Vamsee Jasti}.}
  \bibinfo{year}{2022}\natexlab{}.
\newblock \bibinfo{booktitle}{\emph{Five things we've learned about Web
  Stories}}.
\newblock
\urldef\tempurl%
\url{https://blog.google/web-creators/five-things-weve-learned-about-web-stories/}
\showURL{%
Retrieved May, 2023 from \tempurl}


\bibitem[\protect\citeauthoryear{Jia, Weiss, Biadsy, Macherey, Johnson, Chen,
  and Wu}{Jia et~al\mbox{.}}{2019}]%
        {Translation_Jia2019}
\bibfield{author}{\bibinfo{person}{Ye Jia}, \bibinfo{person}{Ron~J. Weiss},
  \bibinfo{person}{Fadi Biadsy}, \bibinfo{person}{Wolfgang Macherey},
  \bibinfo{person}{Melvin Johnson}, \bibinfo{person}{Zhifeng Chen}, {and}
  \bibinfo{person}{Yonghui Wu}.} \bibinfo{year}{2019}\natexlab{}.
\newblock \bibinfo{title}{Direct speech-to-speech translation with a
  sequence-to-sequence model}.
\newblock
\newblock
\urldef\tempurl%
\url{https://doi.org/10.48550/ARXIV.1904.06037}
\showDOI{\tempurl}


\bibitem[\protect\citeauthoryear{Jun, Bustamante, Whang, and Bischof}{Jun
  et~al\mbox{.}}{2019}]%
        {AMPUp_2019}
\bibfield{author}{\bibinfo{person}{Byungjin Jun},
  \bibinfo{person}{Fabi\'{a}n~E. Bustamante}, \bibinfo{person}{Sung~Yoon
  Whang}, {and} \bibinfo{person}{Zachary~S. Bischof}.}
  \bibinfo{year}{2019}\natexlab{}.
\newblock \showarticletitle{AMP up Your Mobile Web Experience: Characterizing
  the Impact of Google's Accelerated Mobile Project}. In
  \bibinfo{booktitle}{\emph{The 25th Annual International Conference on Mobile
  Computing and Networking}} (Los Cabos, Mexico)
  \emph{(\bibinfo{series}{MobiCom '19})}. \bibinfo{publisher}{Association for
  Computing Machinery}, \bibinfo{address}{New York, NY, USA}, Article
  \bibinfo{articleno}{4}, \bibinfo{numpages}{14}~pages.
\newblock
\showISBNx{9781450361699}
\urldef\tempurl%
\url{https://doi.org/10.1145/3300061.3300137}
\showDOI{\tempurl}


\bibitem[\protect\citeauthoryear{Kalender, Eren, Wu, Cirakman, Kutluk,
  Gultekin, and Korkmaz}{Kalender et~al\mbox{.}}{2018}]%
        {Videolization_Springer2018}
\bibfield{author}{\bibinfo{person}{Murat Kalender}, \bibinfo{person}{M~Tolga
  Eren}, \bibinfo{person}{Zonghuan Wu}, \bibinfo{person}{Ozgun Cirakman},
  \bibinfo{person}{Sezer Kutluk}, \bibinfo{person}{Gunay Gultekin}, {and}
  \bibinfo{person}{Emin~Erkan Korkmaz}.} \bibinfo{year}{2018}\natexlab{}.
\newblock \showarticletitle{Videolization: knowledge graph based automated
  video generation from web content}.
\newblock \bibinfo{journal}{\emph{Multimedia tools and applications}}
  \bibinfo{volume}{77}, \bibinfo{number}{1} (\bibinfo{year}{2018}),
  \bibinfo{pages}{567--595}.
\newblock


\bibitem[\protect\citeauthoryear{Kim, Choi, Kahng, and Kim}{Kim
  et~al\mbox{.}}{2022}]%
        {Kim2022FitVid}
\bibfield{author}{\bibinfo{person}{Jeongyeon Kim}, \bibinfo{person}{Yubin
  Choi}, \bibinfo{person}{Minsuk Kahng}, {and} \bibinfo{person}{Juho Kim}.}
  \bibinfo{year}{2022}\natexlab{}.
\newblock \showarticletitle{FitVid: Responsive and Flexible Video Content
  Adaptation}. In \bibinfo{booktitle}{\emph{Proceedings of the 2022 CHI
  Conference on Human Factors in Computing Systems}} (New Orleans, LA, USA)
  \emph{(\bibinfo{series}{CHI '22})}. \bibinfo{publisher}{Association for
  Computing Machinery}, \bibinfo{address}{New York, NY, USA}, Article
  \bibinfo{articleno}{501}, \bibinfo{numpages}{16}~pages.
\newblock
\showISBNx{9781450391573}
\urldef\tempurl%
\url{https://doi.org/10.1145/3491102.3501948}
\showDOI{\tempurl}


\bibitem[\protect\citeauthoryear{Kim, Guo, Cai, Li, Gajos, and Miller}{Kim
  et~al\mbox{.}}{2014}]%
        {kim2014data}
\bibfield{author}{\bibinfo{person}{Juho Kim}, \bibinfo{person}{Philip~J Guo},
  \bibinfo{person}{Carrie~J Cai}, \bibinfo{person}{Shang-Wen Li},
  \bibinfo{person}{Krzysztof~Z Gajos}, {and} \bibinfo{person}{Robert~C
  Miller}.} \bibinfo{year}{2014}\natexlab{}.
\newblock \showarticletitle{Data-driven interaction techniques for improving
  navigation of educational videos}. In \bibinfo{booktitle}{\emph{Proceedings
  of the 27th annual ACM symposium on User interface software and technology}}.
  \bibinfo{pages}{563--572}.
\newblock


\bibitem[\protect\citeauthoryear{Kupiec, Pedersen, and Chen}{Kupiec
  et~al\mbox{.}}{1995}]%
        {kupiec1995trainable}
\bibfield{author}{\bibinfo{person}{Julian Kupiec}, \bibinfo{person}{Jan
  Pedersen}, {and} \bibinfo{person}{Francine Chen}.}
  \bibinfo{year}{1995}\natexlab{}.
\newblock \showarticletitle{A trainable document summarizer}. In
  \bibinfo{booktitle}{\emph{Proceedings of the 18th annual international ACM
  SIGIR conference on Research and development in information retrieval}}.
  \bibinfo{pages}{68--73}.
\newblock


\bibitem[\protect\citeauthoryear{Kuznetsov, Chang, Hahn, Rachatasumrit,
  Breneisen, Coupland, and Kittur}{Kuznetsov et~al\mbox{.}}{2022}]%
        {fuse_uist22}
\bibfield{author}{\bibinfo{person}{Andrew Kuznetsov},
  \bibinfo{person}{Joseph~Chee Chang}, \bibinfo{person}{Nathan Hahn},
  \bibinfo{person}{Napol Rachatasumrit}, \bibinfo{person}{Bradley Breneisen},
  \bibinfo{person}{Julina Coupland}, {and} \bibinfo{person}{Aniket Kittur}.}
  \bibinfo{year}{2022}\natexlab{}.
\newblock \showarticletitle{Fuse: In-Situ Sensemaking Support in the Browser}.
  In \bibinfo{booktitle}{\emph{Proceedings of the 35th Annual ACM Symposium on
  User Interface Software and Technology}} (Bend, OR, USA)
  \emph{(\bibinfo{series}{UIST '22})}. \bibinfo{publisher}{Association for
  Computing Machinery}, \bibinfo{address}{New York, NY, USA}, Article
  \bibinfo{articleno}{34}, \bibinfo{numpages}{15}~pages.
\newblock
\showISBNx{9781450393201}
\urldef\tempurl%
\url{https://doi.org/10.1145/3526113.3545693}
\showDOI{\tempurl}


\bibitem[\protect\citeauthoryear{Leake, Shin, Kim, and Agrawala}{Leake
  et~al\mbox{.}}{2020}]%
        {TextVisualSlideshows2020}
\bibfield{author}{\bibinfo{person}{Mackenzie Leake},
  \bibinfo{person}{Hijung~Valentina Shin}, \bibinfo{person}{Joy~O. Kim}, {and}
  \bibinfo{person}{Maneesh Agrawala}.} \bibinfo{year}{2020}\natexlab{}.
\newblock \showarticletitle{Generating Audio-Visual Slideshows from Text
  Articles Using Word Concreteness}. In \bibinfo{booktitle}{\emph{Proceedings
  of the 2020 CHI Conference on Human Factors in Computing Systems}} (Honolulu,
  HI, USA) \emph{(\bibinfo{series}{CHI ’20})}.
  \bibinfo{publisher}{Association for Computing Machinery},
  \bibinfo{address}{New York, NY, USA}, \bibinfo{pages}{1–11}.
\newblock
\showISBNx{9781450367080}
\urldef\tempurl%
\url{https://doi.org/10.1145/3313831.3376519}
\showDOI{\tempurl}


\bibitem[\protect\citeauthoryear{Lee, Jiang, Essa, Le, Gong, Yang, and
  Yang}{Lee et~al\mbox{.}}{2020}]%
        {lee2020neural}
\bibfield{author}{\bibinfo{person}{Hsin-Ying Lee}, \bibinfo{person}{Lu Jiang},
  \bibinfo{person}{Irfan Essa}, \bibinfo{person}{Phuong~B Le},
  \bibinfo{person}{Haifeng Gong}, \bibinfo{person}{Ming-Hsuan Yang}, {and}
  \bibinfo{person}{Weilong Yang}.} \bibinfo{year}{2020}\natexlab{}.
\newblock \showarticletitle{Neural design network: Graphic layout generation
  with constraints}. In \bibinfo{booktitle}{\emph{European Conference on
  Computer Vision}}. Springer, \bibinfo{pages}{491--506}.
\newblock


\bibitem[\protect\citeauthoryear{Li, Chen, Tung, and Chilton}{Li
  et~al\mbox{.}}{2021}]%
        {DialogSummarization_UIST21}
\bibfield{author}{\bibinfo{person}{Daniel Li}, \bibinfo{person}{Thomas Chen},
  \bibinfo{person}{Albert Tung}, {and} \bibinfo{person}{Lydia~B Chilton}.}
  \bibinfo{year}{2021}\natexlab{}.
\newblock \showarticletitle{Hierarchical Summarization for Longform Spoken
  Dialog}. In \bibinfo{booktitle}{\emph{The 34th Annual ACM Symposium on User
  Interface Software and Technology}} (Virtual Event, USA)
  \emph{(\bibinfo{series}{UIST '21})}. \bibinfo{publisher}{Association for
  Computing Machinery}, \bibinfo{address}{New York, NY, USA},
  \bibinfo{pages}{582–597}.
\newblock
\showISBNx{9781450386357}
\urldef\tempurl%
\url{https://doi.org/10.1145/3472749.3474771}
\showDOI{\tempurl}


\bibitem[\protect\citeauthoryear{Li, Zhu, Ma, Zhang, and Zong}{Li
  et~al\mbox{.}}{2017}]%
        {li2017multi}
\bibfield{author}{\bibinfo{person}{Haoran Li}, \bibinfo{person}{Junnan Zhu},
  \bibinfo{person}{Cong Ma}, \bibinfo{person}{Jiajun Zhang}, {and}
  \bibinfo{person}{Chengqing Zong}.} \bibinfo{year}{2017}\natexlab{}.
\newblock \showarticletitle{Multi-modal summarization for asynchronous
  collection of text, image, audio and video}. In
  \bibinfo{booktitle}{\emph{Proceedings of the 2017 Conference on Empirical
  Methods in Natural Language Processing}}. \bibinfo{pages}{1092--1102}.
\newblock


\bibitem[\protect\citeauthoryear{Li, Chen, Gao, Chan, Zhao, and Yan}{Li
  et~al\mbox{.}}{2020}]%
        {li2020vmsmo}
\bibfield{author}{\bibinfo{person}{Mingzhe Li}, \bibinfo{person}{Xiuying Chen},
  \bibinfo{person}{Shen Gao}, \bibinfo{person}{Zhangming Chan},
  \bibinfo{person}{Dongyan Zhao}, {and} \bibinfo{person}{Rui Yan}.}
  \bibinfo{year}{2020}\natexlab{}.
\newblock \showarticletitle{VMSMO: Learning to generate multimodal summary for
  video-based news articles}.
\newblock \bibinfo{journal}{\emph{arXiv preprint arXiv:2010.05406}}
  (\bibinfo{year}{2020}).
\newblock


\bibitem[\protect\citeauthoryear{Liu, Saleh, Pot, Goodrich, Sepassi, Kaiser,
  and Shazeer}{Liu et~al\mbox{.}}{2018}]%
        {liu2018generating}
\bibfield{author}{\bibinfo{person}{Peter~J Liu}, \bibinfo{person}{Mohammad
  Saleh}, \bibinfo{person}{Etienne Pot}, \bibinfo{person}{Ben Goodrich},
  \bibinfo{person}{Ryan Sepassi}, \bibinfo{person}{Lukasz Kaiser}, {and}
  \bibinfo{person}{Noam Shazeer}.} \bibinfo{year}{2018}\natexlab{}.
\newblock \showarticletitle{Generating wikipedia by summarizing long
  sequences}.
\newblock \bibinfo{journal}{\emph{arXiv preprint arXiv:1801.10198}}
  (\bibinfo{year}{2018}).
\newblock


\bibitem[\protect\citeauthoryear{Liu, Kirilyuk, Yuan, Olwal, Chi, Chen, and
  Du}{Liu et~al\mbox{.}}{2023}]%
        {VisualCaptions_CHI23}
\bibfield{author}{\bibinfo{person}{Xingyu~"Bruce" Liu},
  \bibinfo{person}{Vladimir Kirilyuk}, \bibinfo{person}{Xiuxiu Yuan},
  \bibinfo{person}{Alex Olwal}, \bibinfo{person}{Peggy Chi},
  \bibinfo{person}{Xiang~"Anthony" Chen}, {and} \bibinfo{person}{Ruofei Du}.}
  \bibinfo{year}{2023}\natexlab{}.
\newblock \showarticletitle{Visual Captions: Augmenting Verbal Communication
  with On-the-Fly Visuals}. In \bibinfo{booktitle}{\emph{Proceedings of the
  2023 CHI Conference on Human Factors in Computing Systems}} (Hamburg,
  Germany) \emph{(\bibinfo{series}{CHI '23})}. \bibinfo{publisher}{Association
  for Computing Machinery}, \bibinfo{address}{New York, NY, USA}, Article
  \bibinfo{articleno}{108}, \bibinfo{numpages}{20}~pages.
\newblock
\showISBNx{9781450394215}
\urldef\tempurl%
\url{https://doi.org/10.1145/3544548.3581566}
\showDOI{\tempurl}


\bibitem[\protect\citeauthoryear{Luhn}{Luhn}{1958}]%
        {luhn1958automatic}
\bibfield{author}{\bibinfo{person}{Hans~Peter Luhn}.}
  \bibinfo{year}{1958}\natexlab{}.
\newblock \showarticletitle{The automatic creation of literature abstracts}.
\newblock \bibinfo{journal}{\emph{IBM Journal of research and development}}
  \bibinfo{volume}{2}, \bibinfo{number}{2} (\bibinfo{year}{1958}),
  \bibinfo{pages}{159--165}.
\newblock


\bibitem[\protect\citeauthoryear{Mani and Bloedorn}{Mani and Bloedorn}{1998}]%
        {mani1998machine}
\bibfield{author}{\bibinfo{person}{Inderjeet Mani} {and} \bibinfo{person}{Eric
  Bloedorn}.} \bibinfo{year}{1998}\natexlab{}.
\newblock \showarticletitle{Machine learning of generic and user-focused
  summarization}. In \bibinfo{booktitle}{\emph{AAAI/IAAI}}.
  \bibinfo{pages}{821--826}.
\newblock


\bibitem[\protect\citeauthoryear{McRoberts, Ma, Hall, and Yarosh}{McRoberts
  et~al\mbox{.}}{2017}]%
        {SnapchatStories_chi17}
\bibfield{author}{\bibinfo{person}{Sarah McRoberts}, \bibinfo{person}{Haiwei
  Ma}, \bibinfo{person}{Andrew Hall}, {and} \bibinfo{person}{Svetlana Yarosh}.}
  \bibinfo{year}{2017}\natexlab{}.
\newblock \showarticletitle{Share First, Save Later: Performance of Self
  through Snapchat Stories}. In \bibinfo{booktitle}{\emph{Proceedings of the
  2017 CHI Conference on Human Factors in Computing Systems}} (Denver,
  Colorado, USA) \emph{(\bibinfo{series}{CHI '17})}.
  \bibinfo{publisher}{Association for Computing Machinery},
  \bibinfo{address}{New York, NY, USA}, \bibinfo{pages}{6902–6911}.
\newblock
\showISBNx{9781450346559}
\urldef\tempurl%
\url{https://doi.org/10.1145/3025453.3025771}
\showDOI{\tempurl}


\bibitem[\protect\citeauthoryear{MediaWiki}{MediaWiki}{2022}]%
        {MediaWiki}
\bibfield{author}{\bibinfo{person}{MediaWiki}.}
  \bibinfo{year}{2022}\natexlab{}.
\newblock \bibinfo{booktitle}{\emph{MediaWiki}}.
\newblock
\urldef\tempurl%
\url{https://www.mediawiki.org/wiki/MediaWiki}
\showURL{%
Retrieved April, 2022 from \tempurl}


\bibitem[\protect\citeauthoryear{Mei and Zhai}{Mei and Zhai}{2008}]%
        {SciLitSummarization_ACLHLT_2008}
\bibfield{author}{\bibinfo{person}{Qiaozhu Mei} {and}
  \bibinfo{person}{ChengXiang Zhai}.} \bibinfo{year}{2008}\natexlab{}.
\newblock \showarticletitle{Generating impact-based summaries for scientific
  literature}. In \bibinfo{booktitle}{\emph{Proceedings of ACL-08: HLT}}.
  \bibinfo{pages}{816--824}.
\newblock


\bibitem[\protect\citeauthoryear{Modani, Srinivasan, and Jhamtani}{Modani
  et~al\mbox{.}}{2016}]%
        {modani2016generating}
\bibfield{author}{\bibinfo{person}{Natwar Modani},
  \bibinfo{person}{Balaji~Vasan Srinivasan}, {and} \bibinfo{person}{Harsh
  Jhamtani}.} \bibinfo{year}{2016}\natexlab{}.
\newblock \showarticletitle{Generating multiple diverse summaries}. In
  \bibinfo{booktitle}{\emph{International Conference on Web Information Systems
  Engineering}}. Springer, \bibinfo{pages}{190--198}.
\newblock


\bibitem[\protect\citeauthoryear{Mohorovi{\v{c}}i{\'c}}{Mohorovi{\v{c}}i{\'c}}{2013}]%
        {mohorovivcic2013implementing}
\bibfield{author}{\bibinfo{person}{Sanja Mohorovi{\v{c}}i{\'c}}.}
  \bibinfo{year}{2013}\natexlab{}.
\newblock \showarticletitle{Implementing responsive web design for enhanced web
  presence}. In \bibinfo{booktitle}{\emph{2013 36th International Convention on
  Information and Communication Technology, Electronics and Microelectronics
  (MIPRO)}}. IEEE, \bibinfo{pages}{1206--1210}.
\newblock


\bibitem[\protect\citeauthoryear{Morrison, Tversky, and Betrancourt}{Morrison
  et~al\mbox{.}}{2000}]%
        {morrison2000animation}
\bibfield{author}{\bibinfo{person}{Julie~Bauer Morrison},
  \bibinfo{person}{Barbara Tversky}, {and} \bibinfo{person}{Mireille
  Betrancourt}.} \bibinfo{year}{2000}\natexlab{}.
\newblock \showarticletitle{Animation: Does it facilitate learning}. In
  \bibinfo{booktitle}{\emph{AAAI spring symposium on smart graphics}},
  Vol.~\bibinfo{volume}{5359}.
\newblock


\bibitem[\protect\citeauthoryear{Nag~Chowdhury, Razniewski, and
  Weikum}{Nag~Chowdhury et~al\mbox{.}}{2021}]%
        {nag2021sandi}
\bibfield{author}{\bibinfo{person}{Sreyasi Nag~Chowdhury},
  \bibinfo{person}{Simon Razniewski}, {and} \bibinfo{person}{Gerhard Weikum}.}
  \bibinfo{year}{2021}\natexlab{}.
\newblock \showarticletitle{SANDI: story-and-images alignment}. In
  \bibinfo{booktitle}{\emph{16th Conference of the European Chapter of the
  Association for Computational Linguistics}}. ACL, \bibinfo{pages}{989--999}.
\newblock


\bibitem[\protect\citeauthoryear{Nav{\'\i}o-Navarro and
  Gonz{\'a}lez-D{\'\i}ez}{Nav{\'\i}o-Navarro and
  Gonz{\'a}lez-D{\'\i}ez}{2021}]%
        {navio2021guiding}
\bibfield{author}{\bibinfo{person}{Marich{\'e} Nav{\'\i}o-Navarro} {and}
  \bibinfo{person}{Laura Gonz{\'a}lez-D{\'\i}ez}.}
  \bibinfo{year}{2021}\natexlab{}.
\newblock \showarticletitle{Guiding the Adoption of News Storytelling Design
  Through Ethics: The Use of Stories in Google’s AMP Project}.
\newblock \bibinfo{journal}{\emph{News Media Innovation Reconsidered: Ethics
  and Values in a Creative Reconstruction of Journalism}}
  (\bibinfo{year}{2021}), \bibinfo{pages}{92--104}.
\newblock


\bibitem[\protect\citeauthoryear{Nenkova and Bagga}{Nenkova and Bagga}{2004}]%
        {EmailSummarization_RANLP2003}
\bibfield{author}{\bibinfo{person}{Ani Nenkova} {and} \bibinfo{person}{Amit
  Bagga}.} \bibinfo{year}{2004}\natexlab{}.
\newblock \showarticletitle{Facilitating email thread access by extractive
  summary generation}.
\newblock \bibinfo{journal}{\emph{Recent advances in natural language
  processing III: selected papers from RANLP}}  \bibinfo{volume}{2003}
  (\bibinfo{year}{2004}), \bibinfo{pages}{287--294}.
\newblock


\bibitem[\protect\citeauthoryear{Oh and Shrobe}{Oh and Shrobe}{2008}]%
        {oh2008generating}
\bibfield{author}{\bibinfo{person}{Alice Oh} {and} \bibinfo{person}{Howard
  Shrobe}.} \bibinfo{year}{2008}\natexlab{}.
\newblock \showarticletitle{Generating baseball summaries from multiple
  perspectives by reordering content}. In \bibinfo{booktitle}{\emph{Proceedings
  of the Fifth International Natural Language Generation Conference}}.
  \bibinfo{pages}{173--176}.
\newblock


\bibitem[\protect\citeauthoryear{Pavel, Goldman, Hartmann, and Agrawala}{Pavel
  et~al\mbox{.}}{2015}]%
        {SceneSkim_UIST15}
\bibfield{author}{\bibinfo{person}{Amy Pavel}, \bibinfo{person}{Dan~B.
  Goldman}, \bibinfo{person}{Bj\"{o}rn Hartmann}, {and}
  \bibinfo{person}{Maneesh Agrawala}.} \bibinfo{year}{2015}\natexlab{}.
\newblock \showarticletitle{SceneSkim: Searching and Browsing Movies Using
  Synchronized Captions, Scripts and Plot Summaries}. In
  \bibinfo{booktitle}{\emph{Proceedings of the 28th Annual ACM Symposium on
  User Interface Software \& Technology}} (Charlotte, NC, USA)
  \emph{(\bibinfo{series}{UIST '15})}. \bibinfo{publisher}{Association for
  Computing Machinery}, \bibinfo{address}{New York, NY, USA},
  \bibinfo{pages}{181–190}.
\newblock
\showISBNx{9781450337793}
\urldef\tempurl%
\url{https://doi.org/10.1145/2807442.2807502}
\showDOI{\tempurl}


\bibitem[\protect\citeauthoryear{Peng, Chi, Kannan, Morris, and Essa}{Peng
  et~al\mbox{.}}{2023}]%
        {SlideGestalt_CHI23}
\bibfield{author}{\bibinfo{person}{Yi-Hao Peng}, \bibinfo{person}{Peggy Chi},
  \bibinfo{person}{Anjuli Kannan}, \bibinfo{person}{Meredith~Ringel Morris},
  {and} \bibinfo{person}{Irfan Essa}.} \bibinfo{year}{2023}\natexlab{}.
\newblock \showarticletitle{Slide Gestalt: Automatic Structure Extraction in
  Slide Decks for Non-Visual Access}. In \bibinfo{booktitle}{\emph{Proceedings
  of the 2023 CHI Conference on Human Factors in Computing Systems}} (Hamburg,
  Germany) \emph{(\bibinfo{series}{CHI '23})}. \bibinfo{publisher}{Association
  for Computing Machinery}, \bibinfo{address}{New York, NY, USA}, Article
  \bibinfo{articleno}{829}, \bibinfo{numpages}{14}~pages.
\newblock
\showISBNx{9781450394215}
\urldef\tempurl%
\url{https://doi.org/10.1145/3544548.3580921}
\showDOI{\tempurl}


\bibitem[\protect\citeauthoryear{Rothe, Narayan, and Severyn}{Rothe
  et~al\mbox{.}}{2020}]%
        {rothe2020leveraging}
\bibfield{author}{\bibinfo{person}{Sascha Rothe}, \bibinfo{person}{Shashi
  Narayan}, {and} \bibinfo{person}{Aliaksei Severyn}.}
  \bibinfo{year}{2020}\natexlab{}.
\newblock \showarticletitle{Leveraging pre-trained checkpoints for sequence
  generation tasks}.
\newblock \bibinfo{journal}{\emph{Transactions of the Association for
  Computational Linguistics}}  \bibinfo{volume}{8} (\bibinfo{year}{2020}),
  \bibinfo{pages}{264--280}.
\newblock


\bibitem[\protect\citeauthoryear{Smith, Yu, Srivastava, Halfaker, Terveen, and
  Zhu}{Smith et~al\mbox{.}}{2020}]%
        {Wikipedia_CHI20}
\bibfield{author}{\bibinfo{person}{C.~Estelle Smith}, \bibinfo{person}{Bowen
  Yu}, \bibinfo{person}{Anjali Srivastava}, \bibinfo{person}{Aaron Halfaker},
  \bibinfo{person}{Loren Terveen}, {and} \bibinfo{person}{Haiyi Zhu}.}
  \bibinfo{year}{2020}\natexlab{}.
\newblock \bibinfo{booktitle}{\emph{Keeping Community in the Loop:
  Understanding Wikipedia Stakeholder Values for Machine Learning-Based
  Systems}}.
\newblock \bibinfo{publisher}{Association for Computing Machinery},
  \bibinfo{address}{New York, NY, USA}, \bibinfo{pages}{1–14}.
\newblock
\showISBNx{9781450367080}
\urldef\tempurl%
\url{https://doi.org/10.1145/3313831.3376783}
\showURL{%
\tempurl}


\bibitem[\protect\citeauthoryear{Song, Tan, Qin, Lu, and Liu}{Song
  et~al\mbox{.}}{2019}]%
        {song2019mass}
\bibfield{author}{\bibinfo{person}{Kaitao Song}, \bibinfo{person}{Xu Tan},
  \bibinfo{person}{Tao Qin}, \bibinfo{person}{Jianfeng Lu}, {and}
  \bibinfo{person}{Tie-Yan Liu}.} \bibinfo{year}{2019}\natexlab{}.
\newblock \showarticletitle{Mass: Masked sequence to sequence pre-training for
  language generation}.
\newblock \bibinfo{journal}{\emph{arXiv preprint arXiv:1905.02450}}
  (\bibinfo{year}{2019}).
\newblock


\bibitem[\protect\citeauthoryear{Sosa-Tzec}{Sosa-Tzec}{2019}]%
        {IG_story_doc19}
\bibfield{author}{\bibinfo{person}{Omar Sosa-Tzec}.}
  \bibinfo{year}{2019}\natexlab{}.
\newblock \showarticletitle{Design Tensions: Interaction Criticism on
  Instagram's Mobile Interface}. In \bibinfo{booktitle}{\emph{Proceedings of
  the 37th ACM International Conference on the Design of Communication}}
  (Portland, Oregon) \emph{(\bibinfo{series}{SIGDOC '19})}.
  \bibinfo{publisher}{Association for Computing Machinery},
  \bibinfo{address}{New York, NY, USA}, Article \bibinfo{articleno}{10},
  \bibinfo{numpages}{10}~pages.
\newblock
\showISBNx{9781450367905}
\urldef\tempurl%
\url{https://doi.org/10.1145/3328020.3353944}
\showDOI{\tempurl}


\bibitem[\protect\citeauthoryear{Srinivasan, Raman, Chen, Bendersky, and
  Najork}{Srinivasan et~al\mbox{.}}{2021}]%
        {Krishna_SIGIR2021}
\bibfield{author}{\bibinfo{person}{Krishna Srinivasan},
  \bibinfo{person}{Karthik Raman}, \bibinfo{person}{Jiecao Chen},
  \bibinfo{person}{Michael Bendersky}, {and} \bibinfo{person}{Marc Najork}.}
  \bibinfo{year}{2021}\natexlab{}.
\newblock \bibinfo{booktitle}{\emph{WIT: Wikipedia-Based Image Text Dataset for
  Multimodal Multilingual Machine Learning}}.
\newblock \bibinfo{publisher}{Association for Computing Machinery},
  \bibinfo{address}{New York, NY, USA}, \bibinfo{pages}{2443–2449}.
\newblock
\showISBNx{9781450380379}
\urldef\tempurl%
\url{https://doi.org/10.1145/3404835.3463257}
\showURL{%
\tempurl}


\bibitem[\protect\citeauthoryear{Tang, Liao, Wang, Sung, Cao, and Lin}{Tang
  et~al\mbox{.}}{2020}]%
        {tang2020supporting}
\bibfield{author}{\bibinfo{person}{Chien-Lin Tang}, \bibinfo{person}{Jingxian
  Liao}, \bibinfo{person}{Hao-Chuan Wang}, \bibinfo{person}{Ching-Ying Sung},
  \bibinfo{person}{Yu-Rong Cao}, {and} \bibinfo{person}{Wen-Chieh Lin}.}
  \bibinfo{year}{2020}\natexlab{}.
\newblock \showarticletitle{Supporting online video learning with concept
  map-based recommendation of learning path}. In
  \bibinfo{booktitle}{\emph{Extended Abstracts of the 2020 CHI Conference on
  Human Factors in Computing Systems}}. \bibinfo{pages}{1--8}.
\newblock


\bibitem[\protect\citeauthoryear{Truong, Chi, Salesin, Essa, and
  Agrawala}{Truong et~al\mbox{.}}{2021}]%
        {MakeupBreakdown_CHI21}
\bibfield{author}{\bibinfo{person}{Anh Truong}, \bibinfo{person}{Peggy Chi},
  \bibinfo{person}{David Salesin}, \bibinfo{person}{Irfan Essa}, {and}
  \bibinfo{person}{Maneesh Agrawala}.} \bibinfo{year}{2021}\natexlab{}.
\newblock \showarticletitle{Automatic Generation of Two-Level Hierarchical
  Tutorials from Instructional Makeup Videos}. In
  \bibinfo{booktitle}{\emph{Proceedings of the 2021 ACM Conference on Human
  Factors in Computing Systems}} \emph{(\bibinfo{series}{CHI '21})}.
\newblock


\bibitem[\protect\citeauthoryear{Turpin, Tsegay, Hawking, and Williams}{Turpin
  et~al\mbox{.}}{2007}]%
        {SearchSummarization_SIGIR2007}
\bibfield{author}{\bibinfo{person}{Andrew Turpin}, \bibinfo{person}{Yohannes
  Tsegay}, \bibinfo{person}{David Hawking}, {and} \bibinfo{person}{Hugh~E
  Williams}.} \bibinfo{year}{2007}\natexlab{}.
\newblock \showarticletitle{Fast generation of result snippets in web search}.
  In \bibinfo{booktitle}{\emph{Proceedings of the 30th annual international ACM
  SIGIR conference on Research and development in information retrieval}}.
  \bibinfo{pages}{127--134}.
\newblock


\bibitem[\protect\citeauthoryear{UzZaman, Bigham, and Allen}{UzZaman
  et~al\mbox{.}}{2011}]%
        {uzzaman2011multimodal}
\bibfield{author}{\bibinfo{person}{Naushad UzZaman}, \bibinfo{person}{Jeffrey~P
  Bigham}, {and} \bibinfo{person}{James~F Allen}.}
  \bibinfo{year}{2011}\natexlab{}.
\newblock \showarticletitle{Multimodal summarization of complex sentences}. In
  \bibinfo{booktitle}{\emph{Proceedings of the 16th international conference on
  Intelligent user interfaces}}. \bibinfo{pages}{43--52}.
\newblock


\bibitem[\protect\citeauthoryear{Vincent, Johnson, and Hecht}{Vincent
  et~al\mbox{.}}{2018}]%
        {Wikipedia_CHI18}
\bibfield{author}{\bibinfo{person}{Nicholas Vincent}, \bibinfo{person}{Isaac
  Johnson}, {and} \bibinfo{person}{Brent Hecht}.}
  \bibinfo{year}{2018}\natexlab{}.
\newblock \showarticletitle{Examining Wikipedia With a Broader Lens:
  Quantifying the Value of Wikipedia's Relationships with Other Large-Scale
  Online Communities}. In \bibinfo{booktitle}{\emph{Proceedings of the 2018 CHI
  Conference on Human Factors in Computing Systems}} (Montreal QC, Canada)
  \emph{(\bibinfo{series}{CHI '18})}. \bibinfo{publisher}{Association for
  Computing Machinery}, \bibinfo{address}{New York, NY, USA},
  \bibinfo{pages}{1–13}.
\newblock
\showISBNx{9781450356206}
\urldef\tempurl%
\url{https://doi.org/10.1145/3173574.3174140}
\showDOI{\tempurl}


\bibitem[\protect\citeauthoryear{Wang, Yang, Hu, Yau, and Shamir}{Wang
  et~al\mbox{.}}{2019}]%
        {WriteAVideo_SIGGRAPH2019}
\bibfield{author}{\bibinfo{person}{Miao Wang}, \bibinfo{person}{Guo-Wei Yang},
  \bibinfo{person}{Shi-Min Hu}, \bibinfo{person}{Shing-Tung Yau}, {and}
  \bibinfo{person}{Ariel Shamir}.} \bibinfo{year}{2019}\natexlab{}.
\newblock \showarticletitle{Write-a-Video: Computational Video Montage from
  Themed Text}.
\newblock \bibinfo{journal}{\emph{ACM Trans. Graph.}} \bibinfo{volume}{38},
  \bibinfo{number}{6}, Article \bibinfo{articleno}{177} (\bibinfo{date}{Nov.}
  \bibinfo{year}{2019}), \bibinfo{numpages}{13}~pages.
\newblock
\showISSN{0730-0301}
\urldef\tempurl%
\url{https://doi.org/10.1145/3355089.3356520}
\showDOI{\tempurl}


\bibitem[\protect\citeauthoryear{Wang, Saharia, Montgomery, Pont-Tuset, Noy,
  Pellegrini, Onoe, Laszlo, Fleet, Soricut, Baldridge, Norouzi, Anderson, and
  Chan}{Wang et~al\mbox{.}}{2023}]%
        {ImagenEditor_CVPR23}
\bibfield{author}{\bibinfo{person}{Su Wang}, \bibinfo{person}{Chitwan Saharia},
  \bibinfo{person}{Ceslee Montgomery}, \bibinfo{person}{Jordi Pont-Tuset},
  \bibinfo{person}{Shai Noy}, \bibinfo{person}{Stefano Pellegrini},
  \bibinfo{person}{Yasumasa Onoe}, \bibinfo{person}{Sarah Laszlo},
  \bibinfo{person}{David Fleet}, \bibinfo{person}{Radu Soricut},
  \bibinfo{person}{Jason Baldridge}, \bibinfo{person}{Mohammad Norouzi},
  \bibinfo{person}{Peter Anderson}, {and} \bibinfo{person}{William Chan}.}
  \bibinfo{year}{2023}\natexlab{}.
\newblock \showarticletitle{Imagen Editor and EditBench: Advancing and
  Evaluating Text-Guided Image Inpainting}. In
  \bibinfo{booktitle}{\emph{CVPR}}.
\newblock


\bibitem[\protect\citeauthoryear{Wikipedia}{Wikipedia}{2022}]%
        {WikipediaFirstSentence}
\bibfield{author}{\bibinfo{person}{Wikipedia}.}
  \bibinfo{year}{2022}\natexlab{}.
\newblock \bibinfo{booktitle}{\emph{Wikipedia:Manual of Style/Lead section}}.
\newblock
\urldef\tempurl%
\url{https://en.wikipedia.org/wiki/Wikipedia:Manual_of_Style/Lead_section#First_sentence}
\showURL{%
Retrieved November, 2022 from \tempurl}


\bibitem[\protect\citeauthoryear{Wikipedia}{Wikipedia}{2023}]%
        {WikiStats}
\bibfield{author}{\bibinfo{person}{Wikipedia}.}
  \bibinfo{year}{2023}\natexlab{}.
\newblock \bibinfo{booktitle}{\emph{Wikipedia:Statistics}}.
\newblock
\urldef\tempurl%
\url{https://en.wikipedia.org/wiki/Wikipedia:Statistics}
\showURL{%
Retrieved May, 2023 from \tempurl}


\bibitem[\protect\citeauthoryear{Witbrock and Mittal}{Witbrock and
  Mittal}{1999}]%
        {witbrock1999ultra}
\bibfield{author}{\bibinfo{person}{Michael~J Witbrock} {and}
  \bibinfo{person}{Vibhu~O Mittal}.} \bibinfo{year}{1999}\natexlab{}.
\newblock \showarticletitle{Ultra-summarization (poster abstract) a statistical
  approach to generating highly condensed non-extractive summaries}. In
  \bibinfo{booktitle}{\emph{Proceedings of the 22nd annual international ACM
  SIGIR conference on Research and development in information retrieval}}.
  \bibinfo{pages}{315--316}.
\newblock


\bibitem[\protect\citeauthoryear{Yadav, Shrivastava, Mohana~Prasad, Arsikere,
  Patil, Kumar, and Deshmukh}{Yadav et~al\mbox{.}}{2015}]%
        {NonLinearNav_IUI2015}
\bibfield{author}{\bibinfo{person}{Kuldeep Yadav}, \bibinfo{person}{Kundan
  Shrivastava}, \bibinfo{person}{S. Mohana~Prasad}, \bibinfo{person}{Harish
  Arsikere}, \bibinfo{person}{Sonal Patil}, \bibinfo{person}{Ranjeet Kumar},
  {and} \bibinfo{person}{Om Deshmukh}.} \bibinfo{year}{2015}\natexlab{}.
\newblock \showarticletitle{Content-Driven Multi-Modal Techniques for
  Non-Linear Video Navigation}. In \bibinfo{booktitle}{\emph{Proceedings of the
  20th International Conference on Intelligent User Interfaces}} (Atlanta,
  Georgia, USA) \emph{(\bibinfo{series}{IUI '15})}.
  \bibinfo{publisher}{Association for Computing Machinery},
  \bibinfo{address}{New York, NY, USA}, \bibinfo{pages}{333–344}.
\newblock
\showISBNx{9781450333061}
\urldef\tempurl%
\url{https://doi.org/10.1145/2678025.2701408}
\showDOI{\tempurl}


\bibitem[\protect\citeauthoryear{Zhang, Zhao, Saleh, and Liu}{Zhang
  et~al\mbox{.}}{2020}]%
        {zhang2020pegasus}
\bibfield{author}{\bibinfo{person}{Jingqing Zhang}, \bibinfo{person}{Yao Zhao},
  \bibinfo{person}{Mohammad Saleh}, {and} \bibinfo{person}{Peter Liu}.}
  \bibinfo{year}{2020}\natexlab{}.
\newblock \showarticletitle{PEGASUS: Pre-training with extracted gap-sentences
  for abstractive summarization}. In \bibinfo{booktitle}{\emph{International
  Conference on Machine Learning}}. PMLR, \bibinfo{pages}{11328--11339}.
\newblock


\bibitem[\protect\citeauthoryear{Zhao, Lin, Luo, Xu, and Wang}{Zhao
  et~al\mbox{.}}{2017}]%
        {VisualNav_MM2017}
\bibfield{author}{\bibinfo{person}{Baoquan Zhao}, \bibinfo{person}{Shujin Lin},
  \bibinfo{person}{Xiaonan Luo}, \bibinfo{person}{Songhua Xu}, {and}
  \bibinfo{person}{Ruomei Wang}.} \bibinfo{year}{2017}\natexlab{}.
\newblock \showarticletitle{A Novel System for Visual Navigation of Educational
  Videos Using Multimodal Cues}. In \bibinfo{booktitle}{\emph{Proceedings of
  the 25th ACM International Conference on Multimedia}} (Mountain View,
  California, USA) \emph{(\bibinfo{series}{MM '17})}.
  \bibinfo{publisher}{Association for Computing Machinery},
  \bibinfo{address}{New York, NY, USA}, \bibinfo{pages}{1680–1688}.
\newblock
\showISBNx{9781450349062}
\urldef\tempurl%
\url{https://doi.org/10.1145/3123266.3123406}
\showDOI{\tempurl}


\bibitem[\protect\citeauthoryear{Zhu, Li, Liu, Zhou, Zhang, and Zong}{Zhu
  et~al\mbox{.}}{2018}]%
        {zhu2018msmo}
\bibfield{author}{\bibinfo{person}{Junnan Zhu}, \bibinfo{person}{Haoran Li},
  \bibinfo{person}{Tianshang Liu}, \bibinfo{person}{Yu Zhou},
  \bibinfo{person}{Jiajun Zhang}, {and} \bibinfo{person}{Chengqing Zong}.}
  \bibinfo{year}{2018}\natexlab{}.
\newblock \showarticletitle{MSMO: Multimodal summarization with multimodal
  output}. In \bibinfo{booktitle}{\emph{Proceedings of the 2018 conference on
  empirical methods in natural language processing}}.
  \bibinfo{pages}{4154--4164}.
\newblock


\end{thebibliography}

\clearpage
\begin{appendices}
\section{Analysis of Published Web Stories}
\label{appendix:existing_stories}

\begin{table}[h!]
\caption{Analysis of 55 published Web Stories of a wide range of topics, including recipe or cooking (40\%), gardening (21.82\%), and travel (16.36\%). A Story has a link named ``Learn more'' (50.9\%) or ``Read full story'' (16.4\%) to its source article.}
% The Story content commonly summarizes the full article while maintaining the structure.
\centerline{\includegraphics[width=.45\textwidth]{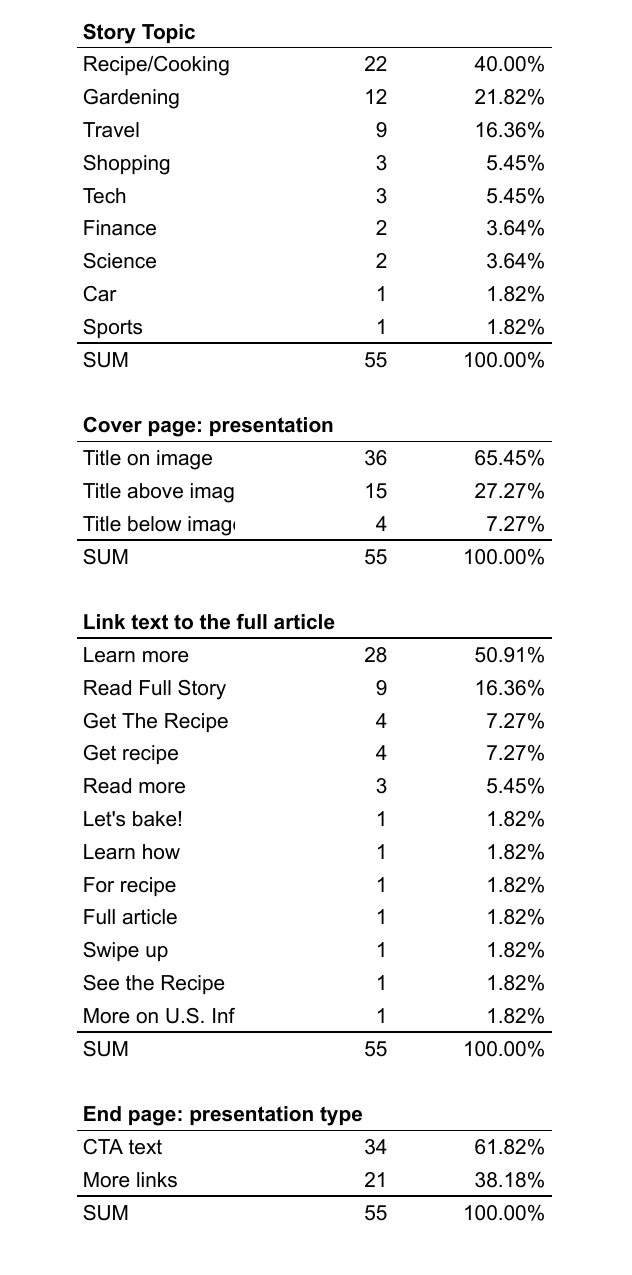}}
\label{tab:online_stories_analysis}
\end{table}
\end{appendices}

\end{document}